\documentclass[12pt]{article}

\usepackage{latexsym, graphicx}
\usepackage{amsmath, amssymb, bm}
\usepackage{hyperref}
\usepackage{longtable}

\newcommand{\ul}[1]{\underline{#1}}
\newcommand{\fb}[1]{\framebox{#1}}

\usepackage{xcolor}

\begin{document}

\begin{titlepage}

\begin{flushright}
USTC-ICTS-14-05\\
\end{flushright}

\begin{center}

\vspace{10mm}

{\LARGE
\textbf{The Statistics of Vacuum Geometry}
}

\vspace{10mm}

{\large
Melissa Duncan\footnote{Email: m.duncan@maths.oxon.org}$^{,a}$,
Wei Gu\footnote{Email: guwei@mail.ustc.edu.cn}$^{,b}$,
Yang-Hui He\footnote{Email: hey@maths.ox.ac.uk}$^{,a,c,d}$,
and Da Zhou\footnote{Email: zhouda@mail.ustc.edu.cn}$^{,b}$\\
}

\vspace{4mm}

{\it
\footnotesize{$^a$ Merton College, University of Oxford, OX1 4JD, UK\\

$^b$ The Interdisciplinary Center for Theoretical Study,\\
University of Science and Technology of China,\\
Hefei, Anhui, 230026, China\\

$^c$ Department of Mathematics, City University, London, EC1V 0HB, UK\\

$^d$ School of Physics, NanKai University, Tianjin, 300071, P.R. China
}
}
\end{center}

\vspace{10mm}

\begin{abstract}
\noindent
We investigate the vacuum moduli space of supersymmetric gauge theories en masse by probing the space of such vacua from a statistical standpoint.
Using quiver gauge theories with ${\cal N}=1$ supersymmetry as a testing ground, we sample over a large number of vacua as algebraic varieties, computing explicitly their dimension, degree and Hilbert series. We study the distribution of these geometrical quantities, and also address the question of how likely it is for the moduli space to be Calabi-Yau. 
\end{abstract}

\end{titlepage}

\tableofcontents

\section{Introduction}

One of the key properties of a quantum field theory is its vacuum structure.
Already, in the famous ``Mexican hat'' of the Higgs potential, the vacuum presents us with non-trivial geometry.
In general, the classical expectation values of scalar fields can parameterize intricate manifolds called the vacuum moduli space (VMS), which are then quantum mechanically corrected.
In field theories with supersymmetry, where there is an abundance of these scalar fields, the VMS generically assumes interesting and complicated forms.

One could exploit this ``vacuum'' geometry, for example, to investigate phenomenological issues; this was a programme launched in \cite{Gray:2005sr,Gray:2006jb,Gray:2008yu,Hanany:2010vu}, where sometimes unexpected geometrical signatures are found in the standard model.
In the context of string theory, realizations and interpretations of the algebraic geometry of the vacuum as the low-energy limit of a compactification or holographic scenario is key to the ``geometrical engineering'' of field theories \cite{Katz:1996fh}.
With the advances in computer algebra and modern algebraic geometry \cite{sing,m2,sage,Gray:2008zs} - of whose powers we will make extensive here - the algorithmic {\it geometry of gauge theories} is now a fruitful enterprise \cite{comp-book}.
Recently, efficient and highly parallelizable methods of numerical algebraic geometry have been applied to studying the VMS \cite{Mehta:2012wk,Hauenstein:2012xs}.

Perhaps the most studied class of supersymmetric gauge theories in four dimensions are the so-called {\it quiver gauge theories}, whose matter content and superpotential can be encoded into a finite graph whose nodes represent factors in a product gauge group and whose arrows, bi-fundamental and adjoint fields. Via Higgsing and addition of flavour nodes, more general types of fields can also be incorporated.
The reason for focusing on these theories over the last two decades is two-fold: (1) the product gauge group structure is clearly an important step toward standard-model building  and (2) the {\it generic} quantum field theory engineered from string theory, especially from holography (AdS/CFT), is of quiver type.

While steady progress has been made to understand the classification of ${\cal N}=2$ four-dimensional gauge theories, particularly of (generalized) quiver type \cite{Cecotti:1992rm, Cecotti:2011rv, Gaiotto:2009we, He:1999xj, Bhardwaj:2013qia}, the perhaps more phenomenologically interesting ${\cal N}=1$ theories presently proliferate wildly beyond control.
Nevertheless, some progress has been made in organizing ${\cal N}=1$ quiver theories by introducing appropriate ``order parameters'', be they block-structures \cite{Benvenuti:2004dw,Hanany:2012mb}, geometric invariants such as the Hilbert series \cite{Benvenuti:2006qr}, or combinatorics of finite directed graphs \cite{Hewlett:2009bx}.

By far the most studied and understood class of super-conformal quiver theories are those whose moduli space is a (non-compact) toric Calabi-Yau variety, notably of complex dimension three \cite{Witten:1993yc,Douglas:1997de,Feng:2000mi}.
Here, the virtues of toric geometry engenders a bipartite-graphic description of the gauge theory \cite{Hanany:2005ve,Franco:2005sm}. From a computational point of view, the moduli spaces are easier to handle in the toric case because the gauge groups are Abelian and the fields are simply complex numbers rather than complex matrices.
Thus, we are working over polynomial rings over the complex numbers and all the technology of algebraic geometry naturally applies.

Bearing the above points in mind - the relevance of the VMS, the efficacy of computational geometry and the ubiquity of quivers - a natural question arises: what is the {\it typical} supersymmetric vacuum?
We are reminded of and motivated by a similar problem in pure geometry.
Whereas algebraic curves are topologically classified by genus, surfaces and higher dimensional varieties thus far defy a complete catalogue.
Even in $\mathbb{C}\mathbb{P}^3$, there are cases not yet known concerning the existence of curves of given genus and degree (cf.~Fig.~18 of IV of \cite{hart}).
We shall thus take a statistical approach, using toric quiver gauge theories as a testing ground, and map out the cartography, with identifiers such as dimension, degree, and Hilbert series, of the space of VMS.
Thence, we will have hint at what the ``typical'' vacuum of a gauge theory might be.

The paper is organized as follows.  In section \ref{s:pre},  we describe in detail the quiver gauge theories of our study. We then discuss how the vacuum space can be expressed as an algebraic variety via calculating relevant gauge invariants in section \ref{s:vac}. An outline of the algorithms used to generate both the quiver theories and relevant statistics is given in section \ref{s:alg}. A selection of our catalogued results are presented for an initial test collection and further generalised to a larger group of non-toric theories.  We illustrate in section \ref{s:compu} the VMS landscape with distributions of vacuum spaces with respect to both degree and dimension. We also collect the Hilbert series data in a vast catalogue for the VMS.
Finally, we conclude with prospects in section \ref{s:conc}.

\section{Preliminaries on Quiver Gauge Theories \label{s:pre}}
We begin with a brief review of our illustrative class of quantum field theories, namely supersymmetric gauge theories which afford quiver description.
Such theories are generically expected to have non-trivial (and often Calabi-Yau) vacuum geometry.

\subsection{Quiver Diagrams \label{s:pre-qd}}
Quivers provide a convenient representation of the particle content of gauge theories that describes D-branes on orbifolds \cite{Douglas:1996sw}. A $\textbf{quiver diagram}$, Q, is a finite directed multigraph where each node represents a compact gauge factor $U(N_{i})$ of the total gauge group $\prod_{i}U(N_{i})$. Nodes are customarily labeled by the index $\textit{i}$ of these gauge factors. The edges represent bi-fundamental fields, those which transform under ($U(N_{a}), \overline{U(N_{b})}$) where $U(N_{a})$ and $U(N_{b})$ are the head and tail gauge groups respectively. We denote these fields by $X^{m}_{ab}$ for fields charged under $\textit{a}$ and $\textit{b}$ respectively whilst multiplicities are indexed by $\textit{m}$. Self-loops, where the head and tail gauge groups coincide, are fields transforming under the usual adjoint representation $Ad_{U(N_{i})}$ that we denote by $\phi^{m}_{i}$.

The gauge invariant terms are constructed by contracting the gauge indices between fields hence they correspond to closed paths in the quiver. A generating set for these invariants is easily obtained from the quiver without adjoints by taking the simple cycles. However, when self-loops are present, additional terms including the adjoint fields must be included among the gauge invariants. Henceforth we shall refer to minimal loops in reference to this expanded set of cycles.

\subsection{Quiver Theories \label{s:pre-qt}}
A $\textbf{quiver theory}$ is a pair $(Q, W)$ where $W$ is the superpotential defining the theory. The superpotential is a formal polynomial in the edges of the quiver where each monomial term is gauge invariant. Hence the superpotential takes the form
\begin{equation}
 W=\sum_{n}a_{n}\mathrm{Tr}(X_{ij}X_{jk}\cdots X_{mi}), \label{Eq1}
\end{equation}
where in general $a_{n}\in \mathbb{C}$.
Allowing for fractional branes one can choose the ranks of the nodes freely and the fields are matrices. In the following we consider gauge groups of the form $\Pi_{i}U(1)$ describing a single brane-probe where all the representations are one dimensional. This introduces a significant redundancy in the number of possible superpotentials for each quiver diagram due to the commutativity of the fields.

We consider $\textbf{Calabi-Yau}$ quivers: these are where the vacuum space is Calabi-Yau. A sufficient condition is that the theory is anomaly-free, a requirement independent of the superpotential. It can be stated as a restriction on the quiver diagram: a quiver theory is Calabi-Yau if all nodes have indegree equal to the outdegree.

A toric quiver theory can be defined as a theory in which each field appears exactly twice in the superpotential in separate terms where the terms have opposite sign. This ``toric condition" was first discussed in \cite{Feng:2000mi} and then used to construct bipartite models in \cite{Hanany:2005ve}.  The condition ensures that the moduli space is a toric variety.  For additional details we refer the reader to the books  \cite{Fulton, Oda} and the physics review paper \cite{Leung:1997tw}. We relax the toric condition in our investigations and consider any polynomials in the gauge invariants as possible superpotentials.

\subsection{The Action \label{s:pre-ta}}
We consider a general gauge theory with $\mathcal{N}=1$ supersymmetry. In the presence of a (compact) gauge group $G$ we have chiral superfields $\Phi_{i}$ transforming under $G$ and a vector superfield $V$ transforming under the Lie algebra of $G$.

The standard action \cite{Wess:1992cp}
 in terms of superspace coordinates is
\begin{equation}
S=\int d^{4}x \left[ \int d^{4}\theta\ \Phi^{\dag}_{i}e^{V}\Phi_{i}+
\frac{1}{4g^{2}} \left( \int d^{2}\theta\ {\rm tr} ( W_{\alpha}W^{\alpha} )+
\int d^{2}\theta\ W(\Phi_i)+h.c. \right) \right], \label{Eq2}
\end{equation}
where $W_{\alpha}=i\overline{D}^{2}e^{-V}D_{\alpha}e^{V}$ is the gauge field strength and $W(\Phi_{i})$ is the superpotential which is a holomorphic function of the chiral superfields. We note that any quadratic terms in the superpotential can be integrated out so we only consider monomials of cubic and higher order.

In the usual way, we calculate the conditions on the classical vacuum state in the Wess-Zumino gauge. It reduces to two types of conditions on the scalar components $\phi_{i}$ of the chiral superfields $\Phi_{i}$. We must satisfy the following:
\begin{equation}
\frac{\partial W(\phi_{i})}{\partial \phi_{i}}=0, \label{Eq3}
\end{equation}
\begin{eqnarray}
{\rm and}\quad \sum_{i}\phi^{\dag}_{i}T^{A}\phi_{i}=0. \label{Eq4}
\end{eqnarray}
These are the F-flatness and D-flatness conditions. The space of solutions to these flatness conditions can be parameterised by an algebraic variety.

\section{Parameterizing the Vacuum \label{s:vac}}
Having introduced our chief class of field theories, we now move on to explicitly describe the mathematical technique of computing the VMS.
The tool we will employ is {\it computational algberaic geometry} of affine varieties.

\subsection{Algebraic Varieties \label{s:vac-av}}
Since we will be viewing vacuum space in terms of algebraic geometry  we briefly introduce some necessary terminology. The fundamental objects of study in algebraic geometry are algebraic sets, that is, sets of solutions to polynomial equations. Let $K$ be an algebraically closed field. We introduce an $\textbf{affine variety}$, $X$, as the locus of points in the affine space $A^{K}$ on which a set of polynomials vanish. So we have
\begin{equation}
 X=\{x\in A^{K}\mid f_{i}(x)=0,  \\\forall i\}. \ \label{Eq5}
\end{equation}
Henceforth we shall set $K=\mathbb{C}$ as relevant to this work. It is well known that we can describe any submanifold of $A^{K}$ with a finite set of polynomials.

The geometric properties of an affine variety can be encoded in its coordinate ring. Associated with $\mathbb{C}^{n}$ we have the polynomial ring $\mathbb{C}[x_{1},\cdots,x_{n}]$. We take the ideal generated by the set of polynomials defining $X$ which we denote by $I(X)$ for brevity. Since $I(X)$ is an ideal we take the quotient $\mathbb{C}[x_{1},\cdots,x_{n}]/I(X)$ as our primary object associated with the variety. Tools of algebraic geometry allow one to extract geometric data from this ring.

For example, the $\textbf{dimension}$ of the space has numerous equivalent definitions including the usual Krull dimension of the quotient ring. The $\textbf{degree}$ is closely related to the number of intersection points of the variety with a straight line but the detailed definition is not required for our analysis.

It is possible to define regular maps $\Phi : X\rightarrow Y \subset \mathbb{C}^{m}$ between varieties which are given by $m$ polynomials on the points of $X$ whose images lie in $Y$. These induce maps on the polynomials rings (often written again using the same symbol) which is actually the pullback $\Phi^{\ast}: \mathbb{C}[x_{1},\cdots,x_{m}]/I(Y)\rightarrow \mathbb{C}[x_{1},\cdots,x_{n}]/I(X)$ defined by $\Phi^{\ast}(f)=f(\Phi)$.

We note given two varieties $X$ and $Y$ we can define a product $X\times Y$ which is also a variety. Let $X\subset \mathbb{C}^{n}$ and $Y\subset \mathbb{C}^{m}$ be varieties defined by the sets of polynomials ${f_{i}(x)}$ and ${g_{i}(x)}$ respectively. We have
\begin{equation}
 X\times Y=\{(x,y)\in \mathbb{C}^{n+m}\mid f_{i}(x)=g_{j}(y)=0 \quad   \forall i,j\}. \label{Eq6}
\end{equation}
The quotient ring associated with this variety is simply
  \begin{equation}
 \mathbb{C}[x_{1},\cdots,x_{n+m}]/(I(X)+I(Y)). \label{Eq7}
\end{equation}

We note some further structure relevant to later discussion. Consider a regular map on a product variety $D: X\times Y\rightarrow \mathbb{C}^{n+m}$ that is a Cartesian product of functions $D=(D_{1},D_{2})$ where $D_{1}: X\rightarrow \mathbb{C}^{n}$ and $D_{2}: X\rightarrow \mathbb{C}^{m}$. The image of this map is clearly the product variety
  \begin{equation}
 \Im(D_{1})\times\Im(D_{2})\subset \mathbb{C}^{n}\times \mathbb{C}^{m}. \label{Eq8}
\end{equation}
Equivalently the coordinate ring map we defined earlier has the domain
\[ \mathbb{C}[x_{1},\cdots,x_{n+m}]/(I(\Im(D_{1}))+\Im(D_{2}))). \]

\subsection{The Vacuum as a Moduli Space \label{s:vac-ms}}
Exploiting the extra gauge invariance of the action Eq.(\ref{Eq2}) allows one to write the space of vacuum solutions as a symplectic quotient and hence an algebraic variety. For a more detailed discussion of the main points we refer the reader to the original paper containing these results \cite{Luty:1995sd}.

We can introduce a less restrictive gauge in which the invariance under the complexified gauge group $G^{c}$ remains. The F-terms are covariant under the imaginary part of the gauge group since the superpotential is a holomorphic function of terms with a gauge index. Hence under this transformation of a solution the F-flatness conditions remained satisfied.

For every F-term solution there is only one D-term solution. Under the imaginary part of the gauge group we can rotate the solution on a D-orbit which contains exactly one solution to the D-terms. Hence we have the fact that the vacuum moduli space is given by the symplectic quotient
 \begin{equation}
 \mathcal{M} = \mathcal{F }~//~ G^{c}, \label{Eq9}
\end{equation}
where $\mathcal{F}$ is the F-term solution manifold.

Now using some standard results we relate this to variety. First we notice $\mathcal{F}$ can be naturally viewed as an affine variety since it corresponds to points at which the polynomial conditions of Eq.(\ref{Eq3}) vanish in $\mathbb{C}^{n}$, where $n$ refers to the number of fields. So we have the polynomial ring $\mathcal{F}\cong\mathbb{C}[\phi_{1},\cdots,\phi_{n}]/\langle\partial_{i}W\rangle$. Lutty and Taylor \cite{Luty:1995sd} highlight the fact that given $\mathcal{F}$ as a variety there is a bijection between Eq.(\ref{Eq8}) and the variety defined by the ring of gauge invariant elements in the ring $\mathcal{F}$.

So taking the minimal generating set of gauge invariants $D=\{r_{1},\cdots,r_{m}\}$ we need the image of this map in $S=\mathbb{C}^{m}$. The vacuum moduli space is
 \begin{equation}
 \mathcal{M} \backsimeq \textrm{Im}(\mathcal{F}\stackrel{D}{\longrightarrow} S). \label{Eq10}
\end{equation}

Given the original ingredients of the gauge invariants or minimal loops of the quiver and F-flatness conditions we can calculate features of the vacuum moduli space (VMS). This technique is sometimes referred to as a version of the $\textbf{Forward Algorithm}$.

We utilise Macaulay2 functions to calculate the ideal of the ring map associated to Eq.(\ref{Eq9}), namely $\ker(S\stackrel{D}\longrightarrow\mathcal{F})$. This gives the ideal in $S$ defining the vacuum space. There are inbuilt functions to calculate the dimension, degree and Hilbert series of the associated variety in the standard sense.

\subsection{Note on Disconnected Quivers \label{s:vac-dq}}
We consider a disconnected quiver diagram with connected components $Q_{j}$ and associated fields $\{\Phi_{j}\}$. The sets of minimal loops and hence gauge invariants can be partitioned into sets associated with each connected component. Thus, a superpotential takes the form $W(\Phi_{i})=\sum_{j}W_{j}(\Phi_{j})$. The algebraic variety is thus a product $\mathcal{F}=\prod_{j}\mathcal{F}_{j}$ and the map $D$ defined by the gauge invariants is a Cartesian product of the maps on each $\mathcal{F}_{j}$.

By our earlier discussion of these types of maps in Eq.(\ref{Eq7}), we know the image of the gauge invariants calculated using Eq.(\ref{Eq9}) will be a product of the images of the $\mathcal{F}_{j}$ in $S_{j}\subset S$. Thus we present results only for connected quiver theories.

\section{Outline of Algorithms \label{s:alg}}
Our strategy is to use the following procedure to generate the
statistics in a systematic manner:
\begin{enumerate}
\item Given a pair of integers $(n,e)$ generate every unique quiver
(excluding disconnected quivers) with $n$ nodes and $e$ edges
that satisfies the Calabi-Yau conditions.
\item For each quiver generate all the possible minimal loops of
the quiver corresponding to the gauge invariants.
\item Find all possible superpotentials $W$ with coefficients taken
from a particular integer set whilst removing the extra redundancy
in the superpotentials.
\item For each superpotential calculate the dimension, degree and Hilbert series corresponding
to the algebraic variety describing the vacuum space $\mathcal{M}$.
\end{enumerate}
In these initial investigations we restrict ourselves to quiver
diagrams with $n\leq 10$ and $e\leq 14$ for which
we can obtain data using standard modern computers. Routines were
tested by checking the results for known quivers with small $(n, e)$.
We outline in further detail the algorithms involved in each of these steps.

\subsection{Generating Quivers \label{s:alg-gq}}
Generating all possible edge sets given $n$ nodes by brute force
methods is computationally intensive. Since we are interested in
Calabi-Yau quivers only, we exploit the fact that the Calabi-Yau
condition implies that the quiver is composed of edge disjoint cycles.
We identify all possible combinations of cycle lengths by generating
all tuples $\{c_{1},...,c_{k}\}$ that satisfy $\sum_{k=1}^n kc_{k}=e$,
where $c_{k}$ is the number of $k-$cycles\footnote{By $k-$cycle,
we mean a cycle of length $k$.}.

However to generate all possible graphs of some cycle combination
from scratch is a non-trivial task. The fact that adding
an extra cycle to a graph with $n=N$ and $e=E$ would result in
a new graph with $n=N$ and $e>E$ can considerably reduce the
amount of computations. That is, if we want to generate graphs with
$n=N$ and $e=E$, we can build them by adding one cycle to graphs
with $n=N$ and $e<E$ which we have already generated.

\begin{figure}[!h]
	\centering
	\includegraphics[width=0.5\textwidth]{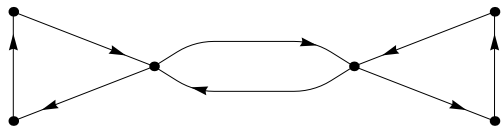}
	\caption{{\sf A Graph of Cycle Combination $\{0,1,2,0,0,0\} $\label{fg:6-8}}}
\end{figure}

\begin{figure}[!h]
	\centering
	\includegraphics[width=0.5\textwidth]{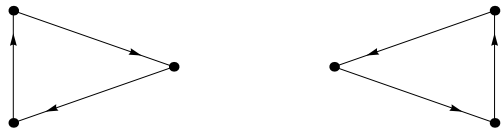}
	\caption{{\sf A Graph of Cycle Combination $\{0,0,2,0,0,0\}$ \label{fg:6-6}}}
\end{figure}

We take $n=6$ and $e=8$ and consider the cycle combination $\{0,1,2,0,0,0\}$ as an example.   A particular graph belonging to this subset is shown in Figure~\ref{fg:6-8}. We can obtain all our desired graphs by simply adding one 2-cycle in every possible way to graphs of the combination $\{0,0,2,0,0,0\}$  which we have generated in a previous process.   For example the Figure~\ref{fg:6-8} graph can be generated by adding a 2-cycle to the graph in Figure~\ref{fg:6-6}. However note in our example that the graph in Figure~\ref{fg:6-6} is disconnected.  When we are interested only in connected graphs, this subtlety means we must continue to reserve disconnected graphs generated in intermediate steps.

Additionally we must be careful when handling graph isomorphisms. It is well known that determining whether two graphs are isomorphic is an NP problem~\cite{Garey:1979}.  With our computational resource limitations we are forced to remove redundant graphs each time we generate the full set of graphs for a given cycle combination. After that any pair of graphs within the same cycle combination are no longer isomorphic.

There is additional redundancy introduced by our algorithm: some graphs can be decomposed into cycle combinations in more than one way.  Thus, when combining the graph sets corresponding to all relevant cycle combinations, we need to remove redundancy once again. The latter is our main obstruction to handling higher $n$ and $e$ values.

Considering only connected quivers reduces the redundant data generated. The vacuum data for all theories can be obtained from these connected cases as discussed in section~\ref{s:vac-dq}.  We note there are more efficient approaches for generating the connected quivers with toric data such as \cite{Hewlett:2009bx}.  For the purposes of this work the more general algorithm was sufficient as the calculation of the vacuum statistics rather than quivers was the limiting factor.

We use the Python-igraph module as a Python language interface for graph manipulations.  This module uses a C library called igraph~\cite{igraph} to perform underlying computations. Thanks to the efficiency of the C language we are able to generate quivers with $n\leq 10$ and $e\leq 14$.

\subsection{Generating Gauge Invariants and Superpotentials \label{s:alg-gio}}
The set of gauge invariants $D$ is found by generating the minimal
loops (see section \ref{s:pre-qd}) of the quiver. For each vertex,
a recursive algorithm traces all possible paths beginning at the
vertex until any vertex in the path is repeated (excluding the cases
when the repeated vertex is a loop). Higher order loops are then
generated by taking all possible combinations of the set of gauge
invariants.

Superpotentials are generated from a set of terms by selecting
coefficients from a coefficient set $\mathcal{C}$. First we take $\mathcal{C}$ to
be $\{-1, 0, 1\}$ as a special case study. Clearly the number
of superpotential terms grows exponentially with the number of
invariants. Given $x$ gauge invariants of three or more fields,
we have $(3^{x}-1)/2$ nontrivial superpotentials. For large numbers of gauge invariants we take random samples of possible superpotentials. In general a sample size s = 500 is taken unless otherwise stated.  We then repeat our analysis for more general coefficients with $\mathcal{C}=\{-5,\cdots,5\}$, where there are now $(11^{x}-1)/2$ non-trivial superpotentials given $x$ invariant terms above quadratic order.

\subsection{Computing the Moduli Space \label{s:alg-ms}}

For each quiver theory generated, the calculations involving algebraic
varieties were performed using the Macaulay2 software~\cite{m2}. We note that
for the purposes of computation the field $\mathbb{C}$ was replaced
with the more manageable $\mathbb{Z}/101\mathbb{Z}$. We found that for
large numbers of superpotential terms (hence F-terms) algorithms became
difficult to run on a standard computer. Fixing the number of superpotential
terms is required to extend analysis to higher order theories.

\subsection{Some Illustrative Examples \label{s:alg-eg}}

\begin{figure}
\begin{center}
\begin{tabular}{cc}
	(1) & (2)  \\
	\raisebox{0pt}[50pt]{\includegraphics[width=0.4\textwidth]{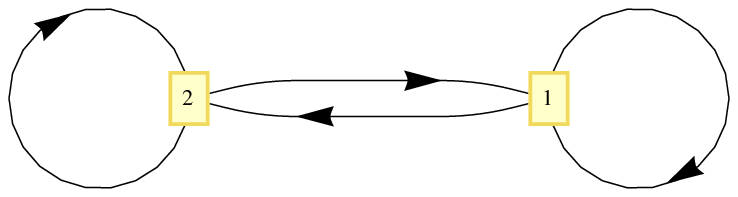}} \\
	\raisebox{-20pt}{(3)}  \\
	\raisebox{10pt}{\includegraphics[width=0.4\textwidth]{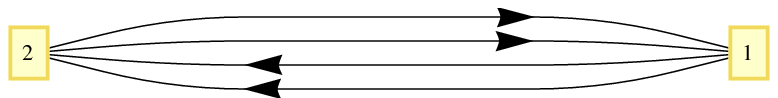}}
	& \raisebox{0pt}[45pt]{\includegraphics[width=0.4\textwidth]{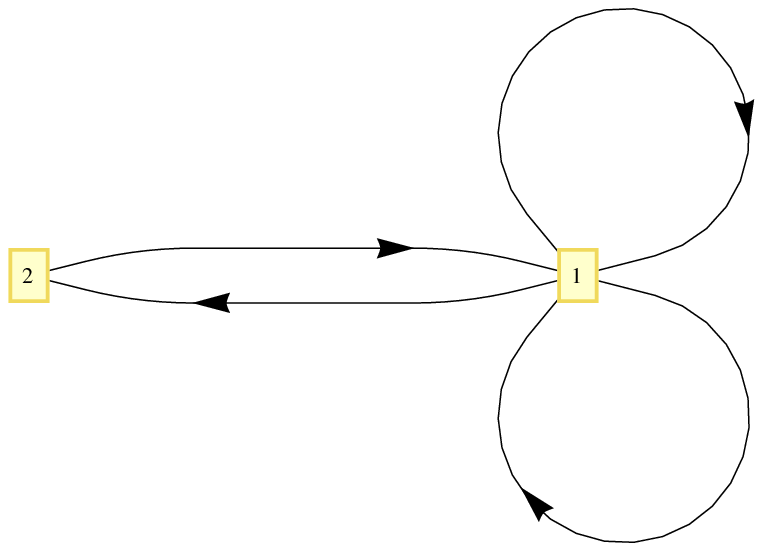}}
\end{tabular}
\end{center}
	\caption{{\sf Unique Connected Quivers for $(n,e)=(2,4)$ \label{fg:il}}}
\end{figure}

As a simple example consider the case of two nodes and four edges. We have three inequivalent quivers presented in Figure~\ref{fg:il}. For each we generate the minimal loops. The first two diagrams have only a single cycle giving the gauge invariant $\{X^{1}_{12}X^{1}_{21}\}$. However including the adjoint fields in each case gives three additional invariants in the generating set. Diagram (3) has a set of four operators $\{X^{1}_{12}X^{1}_{21},\, X^{1}_{12}X^{2}_{21},\, X^{2}_{12}X^{1}_{21},\, X^{2}_{12}X^{2}_{21}\}$ which generate the invariants.  The superscripts of the fields are the indices of the multi-edges.

Considering terms that are first order in the gauge invariants only, we obtain the superpotentials
 \begin{eqnarray}
 W_{1}&=&a_{1}\phi^{1}_{1}X^{1}_{12}X^{1}_{21}+a_{2}\phi^{1}_{2}X^{1}_{12}X^{1}_{21}+a_{3}\phi^{1}_{1}\phi^{1}_{2}X^{1}_{12}X^{1}_{21} \label{Eq11}\\
\text{and}
\quad  W_{2}&=&a_{1}\phi^{1}_{1}X^{1}_{12}X^{1}_{21}+a_{2}\phi^{2}_{1}X^{1}_{12}X^{1}_{21}+a_{3}\phi^{1}_{1}\phi^{2}_{1}X^{1}_{12}X^{1}_{21}, \label{Eq12}
\end{eqnarray}
where $W_{i}$ corresponds to diagram ($i$) and $a_{j}\in\{-1, 0, 1\}$. We note the extra redundancy here due to the commutativity of the fields.

Additionally, since we take all gauge groups to be identical, the position of adjoints in the quiver is only relevant in so far as determining simple cycles the loop can be combined with.  Hence the invariants and VMS for quivers (1) and (2) are identical as the structure does not distinguish between the fields $\phi_{1}$ and $\phi_{2}$.

The VMS details for the first of the pair are listed explicitly in Table~\ref{tb:2n-4e1}.  While the degree of the vacuum space is always one, there are 10 theories with dimension one and 16 of dimension zero.

\begin{table}
\caption{{\sf VMS Data for Quiver (1) of Figure~\ref{fg:il}\label{tb:2n-4e1}}}
\begin{center}
\begin{tabular}{|c|c|c|}
\hline
Superpotential & dim$(\mathcal{M})$ & deg$(\mathcal{M})$\\
\hline
	$\phi_{1}^{1}X_{12}^{1}X_{21}^{1}$
	& 0
	& 1\\
\hline
	$\phi_{2}^{1}X_{12}^{1}X_{21}^{1}$
	& 0
	& 1\\
\hline
	$\phi_{1}^{1}X_{12}^{1}X_{21}^{1}+\phi_{2}^{1}X_{12}^{1}X_{21}^{1}$
	& 0
	& 1\\
\hline
	$-\phi_{1}^{1}X_{12}^{1}X_{21}^{1}+\phi_{2}^{1}X_{12}^{1}X_{21}^{1}$
	& 0
	& 1\\
\hline
	$\phi_{1}^{1}\phi_{2}^{1}X_{12}^{1}X_{21}^{1}$
	& 1
	& 1\\
\hline
    $\phi_{1}^{1}X_{12}^{1}X_{21}^{1}+\phi_{1}^{1}\phi_{2}^{1}X_{12}^{1}X_{21}^{1}$
	& 1
	& 1\\
\hline
    $-\phi_{1}^{1}X_{12}^{1}X_{21}^{1}+\phi_{1}^{1}\phi_{2}^{1}X_{12}^{1}X_{21}^{1}$
	& 1
	& 1\\
\hline
    $\phi_{2}^{1}X_{12}^{1}X_{21}^{1}+\phi_{1}^{1}\phi_{2}^{1}X_{12}^{1}X_{21}^{1}$
	& 1
	& 1\\
\hline
    $\phi_{1}^{1}X_{12}^{1}X_{21}^{1}+\phi_{2}^{1}X_{12}^{1}X_{21}^{1}+\phi_{1}^{1}\phi_{2}^{1}X_{12}^{1}X_{21}^{1}$
	& 0
	& 1\\
\hline
    $-\phi_{1}^{1}X_{12}^{1}X_{21}^{1}+\phi_{2}^{1}X_{12}^{1}X_{21}^{1}+\phi_{1}^{1}\phi_{2}^{1}X_{12}^{1}X_{21}^{1}$
	& 0
	& 1\\
\hline
    $-\phi_{2}^{1}X_{12}^{1}X_{21}^{1}+\phi_{1}^{1}\phi_{2}^{1}X_{12}^{1}X_{21}^{1}$
	& 1
	& 1\\
\hline
    $\phi_{1}^{1}X_{12}^{1}X_{21}^{1}-\phi_{2}^{1}X_{12}^{1}X_{21}^{1}+\phi_{1}^{1}\phi_{2}^{1}X_{12}^{1}X_{21}^{1}$
	& 0
	& 1\\
\hline
    $-\phi_{1}^{1}X_{12}^{1}X_{21}^{1}-\phi_{2}^{1}X_{12}^{1}X_{21}^{1}+\phi_{1}^{1}\phi_{2}^{1}X_{12}^{1}X_{21}^{1}$
	& 0
	& 1\\
\hline
\end{tabular}
\end{center}
\end{table}

Restricting to first order terms in gauge invariants for quiver (3) gives a trivial superpotential as we require monomials of degree three or higher. Including products of invariants to second order we obtain the set of models with
\begin{eqnarray}
 W_{3}&=&a_{1}X^{1}_{12}X^{1}_{21}X^{2}_{12}X^{2}_{21}+a_{2}X^{1}_{12}X^{1}_{21}X^{1}_{12}X^{2}_{21}+a_{3}X^{1}_{12}X^{1}_{21}X^{2}_{21}X^{1}_{12}\nonumber \\ &&+a_{4}X^{1}_{12}X^{1}_{21}X^{1}_{12}X^{1}_{21}
 +a_{5}X^{2}_{12}X^{2}_{21}X^{2}_{12}X^{2}_{21}+a_{6}X^{2}_{21}X^{2}_{12}X^{2}_{21}X^{1}_{12}. \label{Eq13}
\end{eqnarray}

\begin{figure}
	\begin{tabular}{cc}
	\raisebox{-0.1\height}{\includegraphics[width=0.3\textwidth]{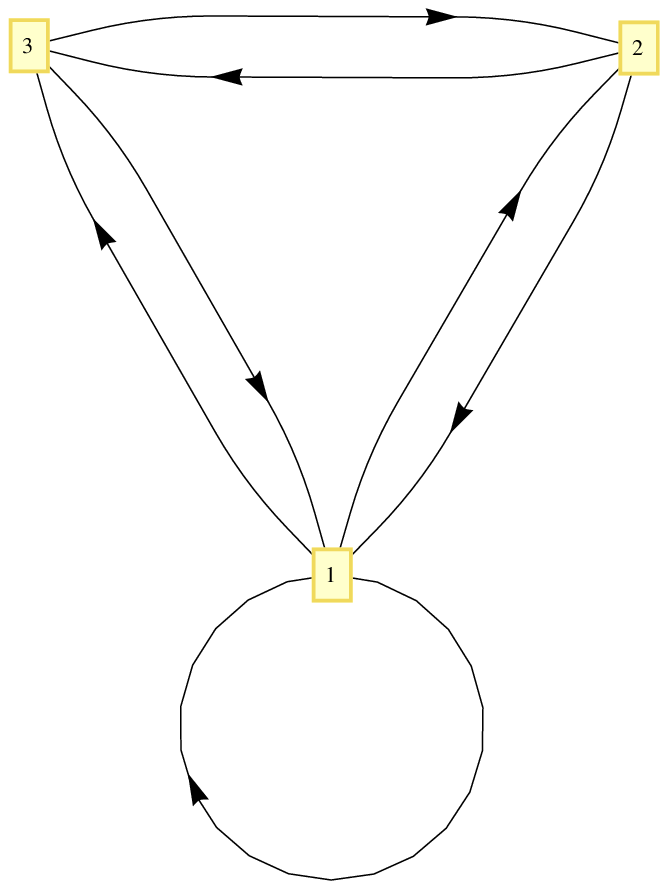}} &
	\includegraphics[width=0.48\textwidth]{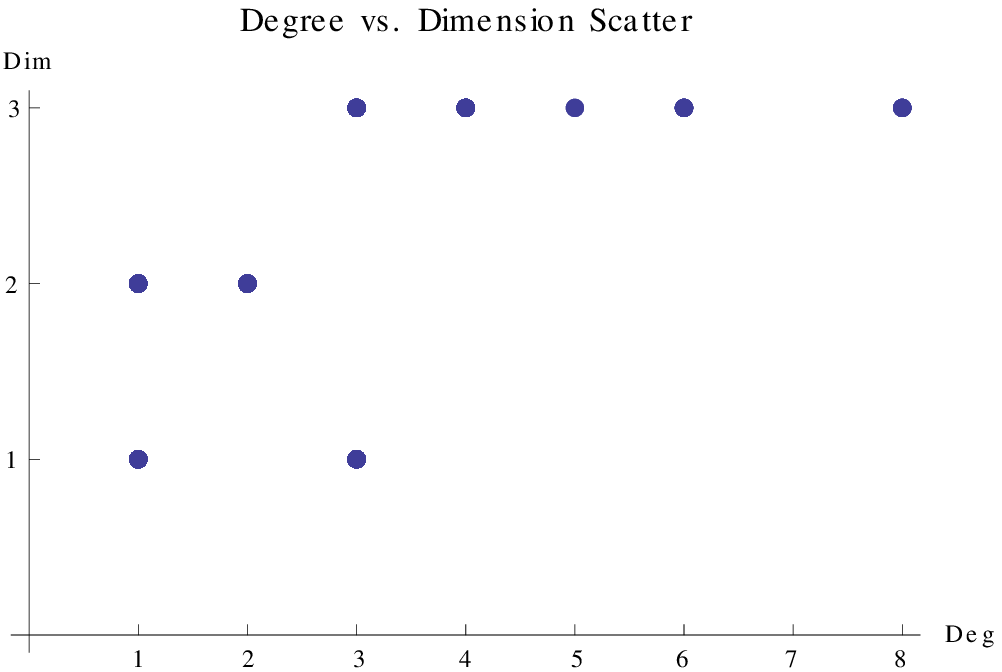} \\
	& \\
	\includegraphics[width=0.48\textwidth]{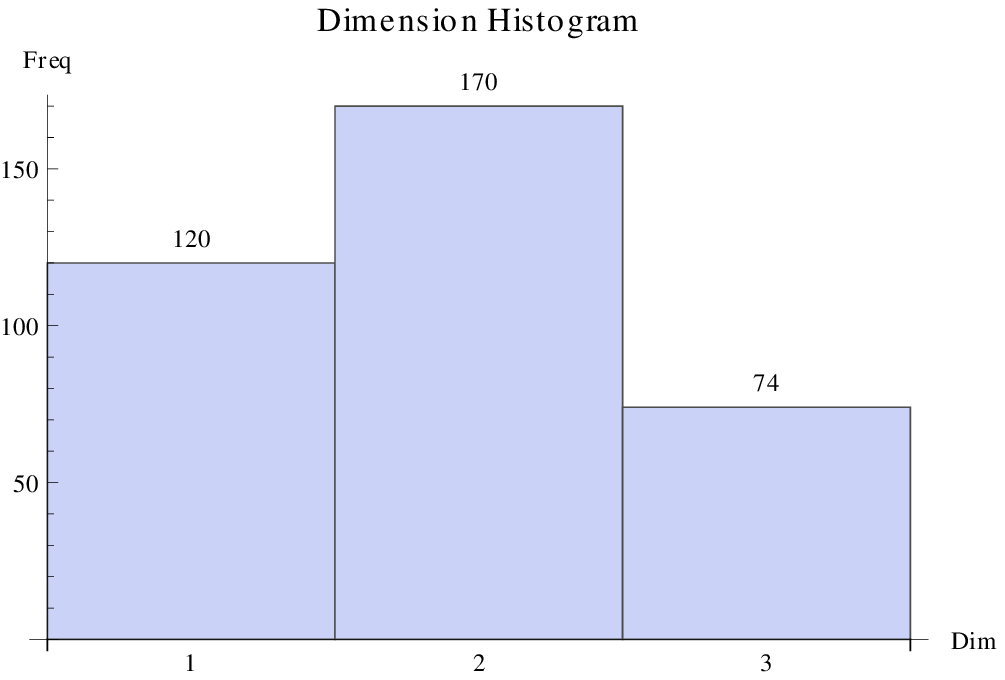} &
	\includegraphics[width=0.48\textwidth]{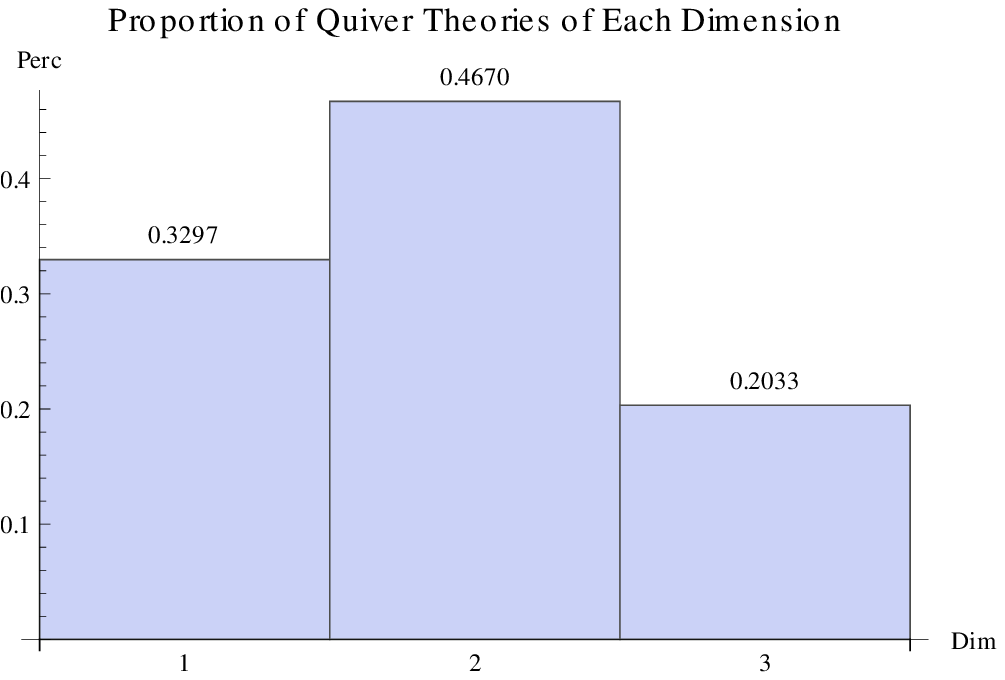} \\
	& \\
	\includegraphics[width=0.48\textwidth]{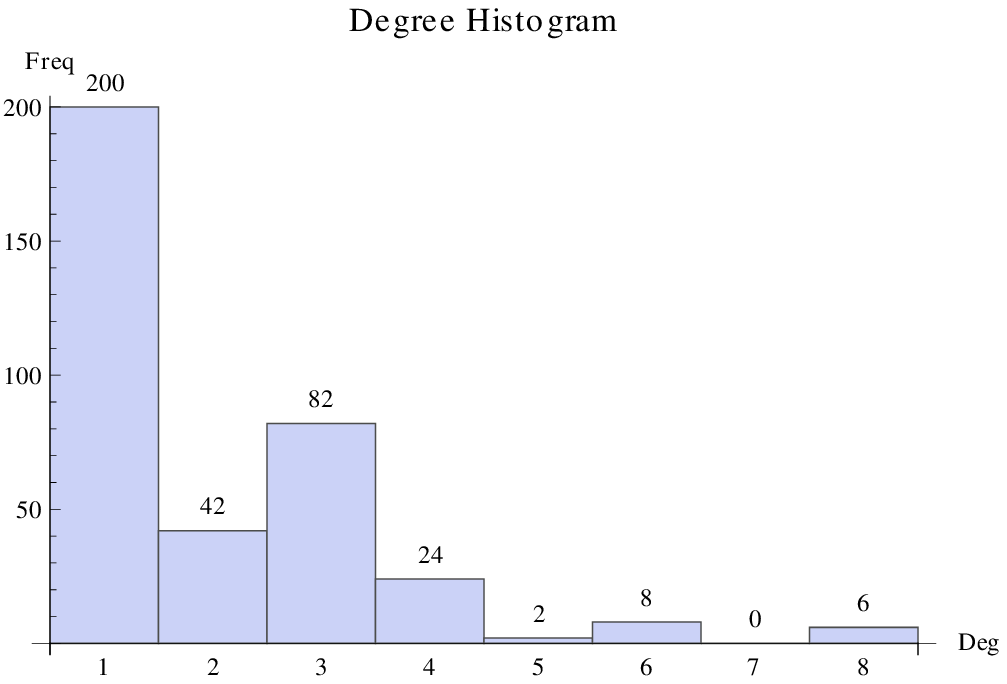} &
	\includegraphics[width=0.48\textwidth]{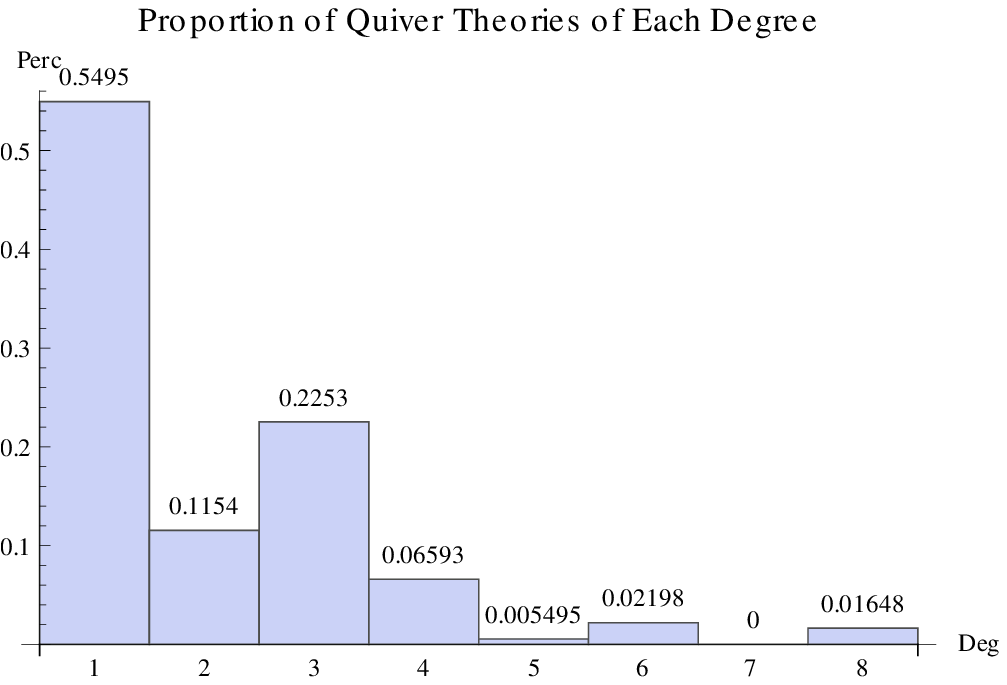} \\
	\end{tabular}
	\caption{{\sf Distribution of Vacuum Spaces for Example Quiver with $(n, e)=(3, 7)$ \label{fg:3-7-10}}}
\end{figure}

Increasing the number of edges rapidly produces theories with large numbers of invariants. Hence we can still obtain interesting results whilst we strict ourselves to superpotentials to first order in the invariants. For example the quiver with $(n, e)=(3, 7)$ shown in Figure~\ref{fg:3-7-10} has the five simple loops
 \begin{equation}
 \{X^{1}_{21}X^{1}_{12},\, X^{1}_{13}X^{1}_{31},\, X^{1}_{23}X^{1}_{32},\, X^{1}_{13}X^{1}_{32}X^{1}_{21},\, X^{1}_{12}X^{1}_{23}X^{1}_{31}\}. \label{Eq14}
\end{equation}
The set of invariants including terms with $\phi^{1}_{1}$ has an additional four terms
\begin{equation}
 \{X^{1}_{31}\phi^{1}_{1}X^{1}_{13},\, X^{1}_{21}\phi^{1}_{1}X^{1}_{12},\, X^{1}_{31}\phi^{1}_{1}X^{1}_{12}X^{1}_{23},\, X^{1}_{32}X^{1}_{21}\phi^{1}_{1}X^{1}_{13}\}. \label{Eq15}
\end{equation}

Taking the full set of gauge invariants and coefficients in the set $\{-1, 0 ,1\}$ we obtain a total of 364 $(x=6)$ non-trivial superpotentials. Details of the VMS for this quiver are given in Figure~\ref{fg:3-7-10}. The majority of theories have a VMS of dimension one, slightly more than double the 74 theories out of 364 that have dimension three.   The distribution of degrees is more interesting; there are more theories with degree one than for degree two up to seven combined.  Notably in this case there are no theories of degree seven.  From the scatter plot we observe there are a limited number of  combinations of degree and dimension occurring in the set of vacuum spaces.

\section{Computational Results \label{s:compu}}

We outline results for quiver theories where superpotentials are
constructed from the terms to first order in the gauge invariants,
that is, terms corresponding to the minimal loops in the quiver
as discussed in section~\ref{s:alg-gio}. Table~\ref{tb:n-e} details
the number of connected quivers generated for the various pairs $(n,e)$.

Entries where the quiver count is underlined or framed indicate we have computed the VMS statistics.  The underlined entries highlight cases for which we considered coefficients in the set $\{-5,\cdots,5\}$ rather than just the simple case study of coefficients in $\{-1, 0 ,1\}$.  We refer to these two cases as the {\bf generic case} and {\bf basic case} respectively from here on. Unmarked quiver counts indicate where we reached the limits of efficient calculation using Macaulay2 either due to the number of terms in the superpotentials or large quiver count.

We present a short selection of the VMS data we generated along with the associated quivers in sections~\ref{s:compu-2n} and \ref{s:compu-3n}. Sample sizes of $s=500$ are used unless otherwise stated. We collate the statistics in general under the pairs $(n, e)$ in section~\ref{s:compu-lds} rather than individual quivers to give a clearer picture of the vacuum landscape. In section~\ref{s:compu-sca} we present a degree versus dimension scatter for all $(n, e)$ we have computed.

\begin{table}[!b]
\caption{\sf Counts of Connected Quivers for Various $(n,e)$.
The underlined and framed cases are those for which we explicit compute all the  relevant geometrical information for the VMS; the framed are at the limit of current computing power.
\label{tb:n-e}}
\begin{center}
\begin{tabular}{c|ccccccccccccc}
n&\multicolumn{13}{c}{e} \\
& 2 & 3 & 4 & 5 & 6 & 7 & 8 & 9 & 10 & 11 & 12 & 13 & 14\\
\hline
2 &1 & \ul{1} & \ul{3} & \ul{3} & \fb{6} & 6 & 10 & 10 & 15 & 15 & 21 & 21 & 28\\
3 &0 & \ul{1} & \ul{2} & \ul{5} & \ul{14} & \fb{24} & 46 & 81 & 130 & 202 & 314 & 452 & 652\\
4 &0 & 0 & \ul{1} & \ul{2} & \ul{12} & \ul{34} & \fb{105} & 245 & 578 & 1201 & 2463 & 4658 & 8658\\
5 &0 & 0 & 0 & \ul{1} & \ul{3} & \ul{18} & \ul{88} & \fb{327} & 1088 & 3187 & 8694 & 22027 & 52944\\
6 &0 & 0 & 0 & 0 & \ul{1} & \ul{3} & \ul{32} & \ul{187} & \fb{942} & 3899 & 14670 & 49515 & 156107\\
7 &0 & 0 & 0 & 0 & 0 & \ul{1} & \ul{4} & \ul{45} & \fb{370} & 2309 & 12224 & 56292 & 234463\\
8 &0 & 0 & 0 & 0 & 0 & 0 & \ul{1} & \ul{4} & \ul{68} & \fb{662} & 5243 & 33654 &189116\\
9 &0 & 0 & 0 & 0 & 0 & 0 & 0 & \ul{1} & \ul{5} & \ul{90} & \fb{1147} & 10804 & 84297\\
10&0 & 0 & 0 & 0 & 0 & 0 & 0 & 0 & \ul{1} & \ul{5} & \ul{126} & \fb{1839} & 21034 \\
\end{tabular}
\end{center}
\end{table}

\subsection{Two Node Quivers \label{s:compu-2n}}

We list some low order examples of quivers with $n=2$ along with some
features in Tables~\ref{tb:2n-5e} and \ref{tb:2n-6e}. The first column
contains the unique quiver diagrams with the given $n$ and $e$. The second indicates whether
random samples are used: ``YES" in this column indicates a sample of size $s=500$.
The third column tables give the distribution of the vacuum spaces according to the dimension or degree of the moduli space.

\begin{table}[!b]
\caption{\sf VMS Statistics for $(2,5)$ Quivers for the Basic Case:
The first two quivers have the same physical content when coefficients are restricted to $\{-1,0,1\}$. Differences in this table are a result of sampling. \label{tb:2n-5e}}
\begin{center}
\begin{tabular}{|c|c|c|c|}
\hline
	QUIVER & SAMPLING & Dim $|$ Count & Deg $|$ Count\\
\hline
	\raisebox{-0.4\height}{\includegraphics[width=4cm]{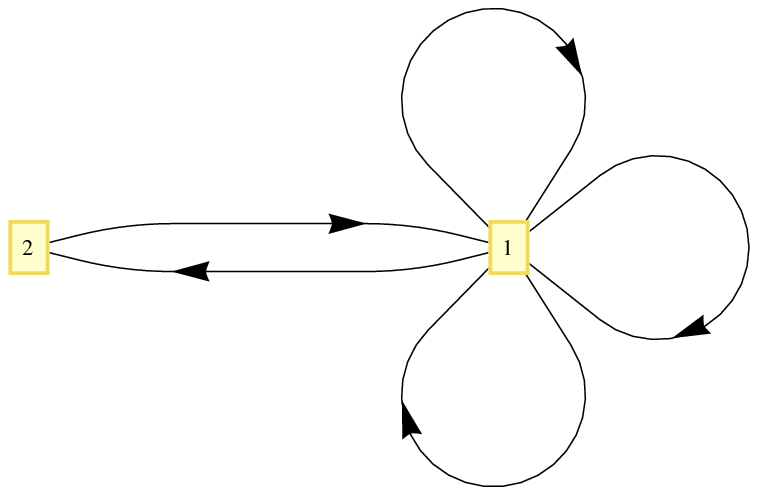}}
	& YES
	& \begin{tabular}{c|c} 0 & 397\\ 1 & 42\\ 2 & 61 \end{tabular}
	& \begin{tabular}{c|c} 1 & 445\\ 2 & 43\\ 3 & 12 \end{tabular}\\
\hline
	\raisebox{-0.4\height}{\includegraphics[width=4cm]{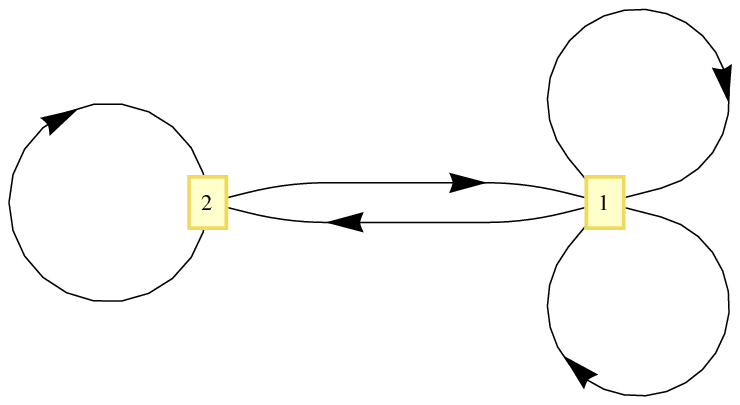}}
	& YES
	& \begin{tabular}{c|c} 0 & 385\\ 1 & 44\\ 2 & 71 \end{tabular}
	& \begin{tabular}{c|c} 1 & 436\\ 2 & 56\\ 3 & 8 \end{tabular}\\
\hline
	\raisebox{-0.4\height}{\includegraphics[width=4cm]{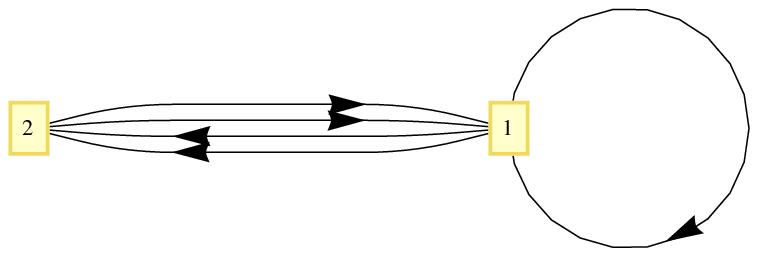}}
	& NO
	& \begin{tabular}{c|c} 2 & 40 \end{tabular}
	& \begin{tabular}{c|c} 2 & 24\\ 3 & 16 \end{tabular}\\
\hline
\end{tabular}
\end{center}
\end{table}

Shown in Table~\ref{tb:2n-5e} are three inequivalent quiver diagrams with $n=2$ and $e=5$. The majority of
theories associated with the first two diagrams have a VMS with dimension zero while about 20 percent have non-vanishing dimension.  A similar distribution occurs for the degree of the VMS for these quivers.  The third quiver has only 40 theories all of dimension two, while the degree of the VMS was either two or three split in the ratio 2:3.

\begin{table}
\caption{\sf VMS Statistics for $(2,6)$ Quivers in Basic Case\label{tb:2n-6e}}
\begin{center}
\begin{tabular}{|c|c|c|c|}
\hline
	QUIVER & SAMPLING & Dim $|$ Count & Deg $|$ Count\\
\hline
	\raisebox{-0.4\height}{\includegraphics[width=4cm]{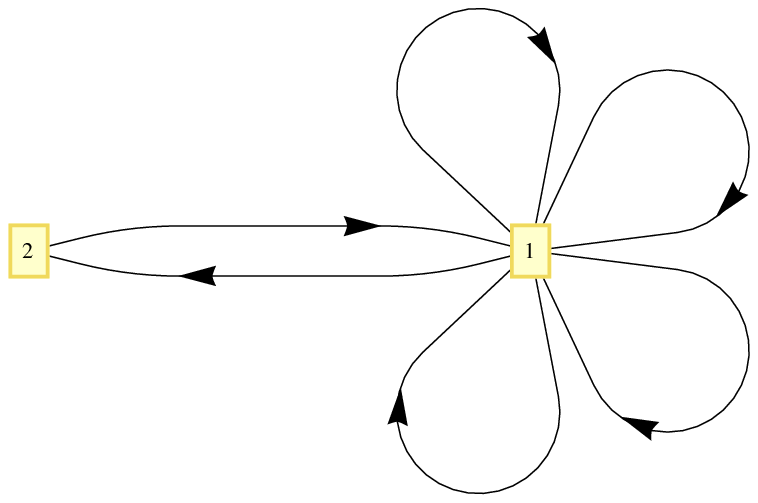}}
	& YES
	& \begin{tabular}{c|c} 0 & 373\\ 1 & 123\\ 2 & 2\\ 3 & 2 \end{tabular}
	& \begin{tabular}{c|c} 1 & 420\\ 2 & 67\\ 3 & 9\\ 4 & 2\\ 6 & 2 \end{tabular}\\
\hline
	\raisebox{-0.4\height}{\includegraphics[width=4cm]{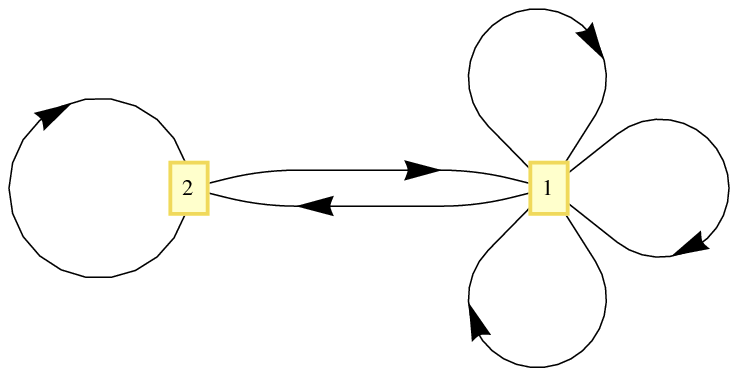}}
	& YES
	& \begin{tabular}{c|c} 0 & 336\\ 1 & 153\\ 2 & 9\\ 3 & 2 \end{tabular}
	& \begin{tabular}{c|c} 1 & 395\\ 2 & 83\\ 3 & 13\\ 4 & 6\\ 5 & 1\\ 6 & 1\\ 10 & 1 \end{tabular}\\
\hline
	\raisebox{-0.4\height}{\includegraphics[width=4cm]{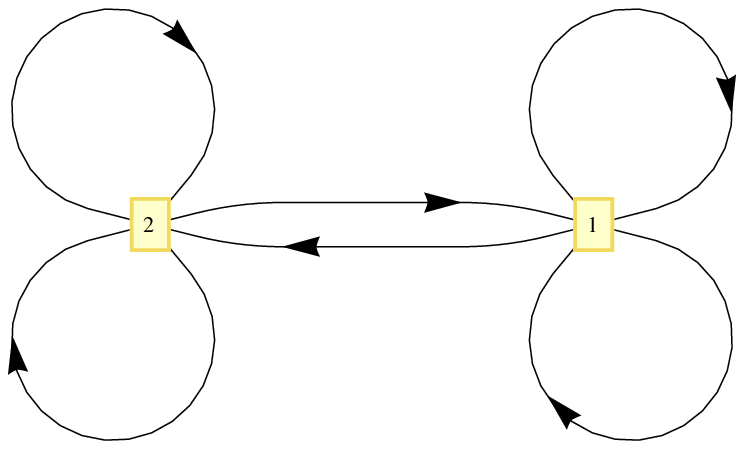}}
	& YES
	& \begin{tabular}{c|c} 0 & 365\\1 & 126\\ 2 & 8\\ 3 & 1 \end{tabular}
	& \begin{tabular}{c|c} 1 & 424\\ 2 & 59\\ 3 & 13\\ 4 & 3\\ 6 & 1 \end{tabular}\\
\hline
	\raisebox{-0.4\height}{\includegraphics[width=4cm]{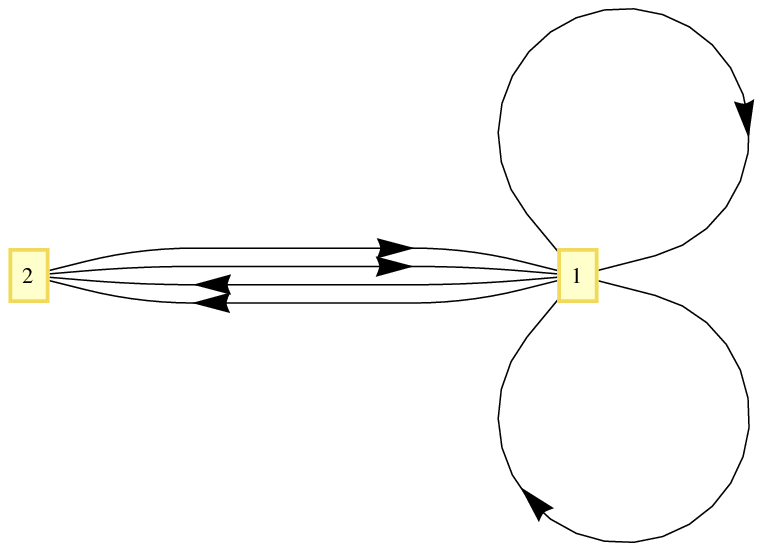}}
	& YES
	& \begin{tabular}{c|c} 1 & 432\\ 2 & 52\\ 3 & 16 \end{tabular}
	& \begin{tabular}{c|c} 1 & 85\\ 2 & 222\\ 3 & 120\\ 4 & 46\\ 5 & 10\\ 6 & 16\\ 10 & 1 \end{tabular}\\
\hline
	\raisebox{-0.4\height}{\includegraphics[width=4cm]{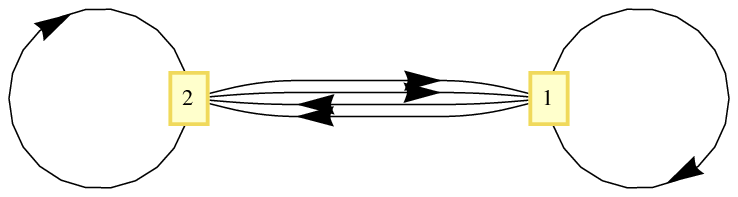}}
	& YES
	& \begin{tabular}{c|c} 1 & 425\\ 2 & 57\\ 3 & 18 \end{tabular}
	& \begin{tabular}{c|c} 1 & 86\\ 2 & 195\\ 3 & 126\\ 4 & 61\\ 5 & 12\\ 6 & 18\\ 10 & 2 \end{tabular}\\
\hline
	\raisebox{-0.4\height}{\includegraphics[width=4cm]{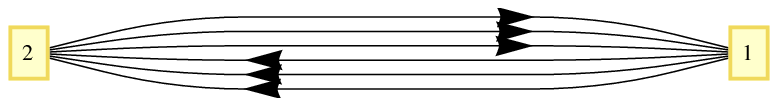}}
	& NO
	& \begin{tabular}{c} N/A \end{tabular}
    & \begin{tabular}{c} N/A \end{tabular}\\
\hline
\end{tabular}
\end{center}
\end{table}

There are six inequivalent quiver diagrams in Table~\ref{tb:2n-6e} when we have added an additional edge. The first three quivers should have the same features since we are only considering the Abelian gauge theory and the quivers only differ in the location of the self-adjoint fields. Similarly, the fourth and fifth quivers should generate the same theories. The small differences in the results for these quivers are due to statistical error. 

Again we see that the VMS of most theories for the first three quivers have dimension zero, while about 25\%--30\% have dimension one and a handful of cases have dimension two or three.  The degree distributions are similar though the degree varies over a larger range of values. The features of the final two quivers are a contrast.  There are no spaces of dimension zero here but rather most theories have dimension one. And the majority of theories have degree two as opposed to degree one. There are no theories subject to our conditions for the last quiver in the table.

\begin{table}[!b]
\caption{\sf VMS Statistics for $(3,3)$ Quivers in Basic Case \label{tb:3n-3e}}
\begin{center}
\begin{tabular}{|c|c|c|c|}
\hline
	QUIVER & SAMPLING & Dim $|$ Count & Deg $|$ Count\\
\hline
	\raisebox{-0.4\height}{\includegraphics[width=4cm]{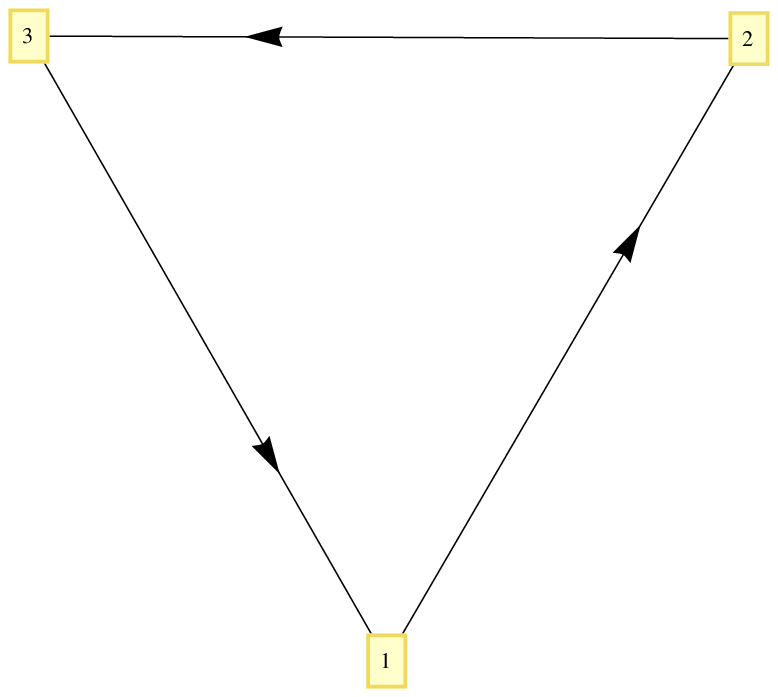}}
	& NO
	& \begin{tabular}{c|c} 0 & 1\end{tabular}
	& \begin{tabular}{c|c} 1 & 1 \end{tabular}\\
\hline
\end{tabular}
\end{center}
\end{table}

\begin{table}[!b]
\caption{\sf VMS Statistics for $(3,4)$ Quivers in the Basic Case \label{tb:3n-4e}}
\begin{center}
\begin{tabular}{|c|c|c|c|}
\hline
	QUIVER & SAMPLING & Dim $|$ Count & Deg $|$ Count\\
\hline
	\raisebox{-0.4\height}{\includegraphics[width=4cm]{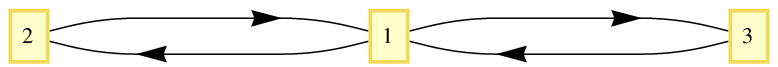}}
	& NO
	& \begin{tabular}{c} N/A \end{tabular}
	& \begin{tabular}{c} N/A  \end{tabular}\\
\hline
	\raisebox{-0.4\height}{\includegraphics[width=4cm]{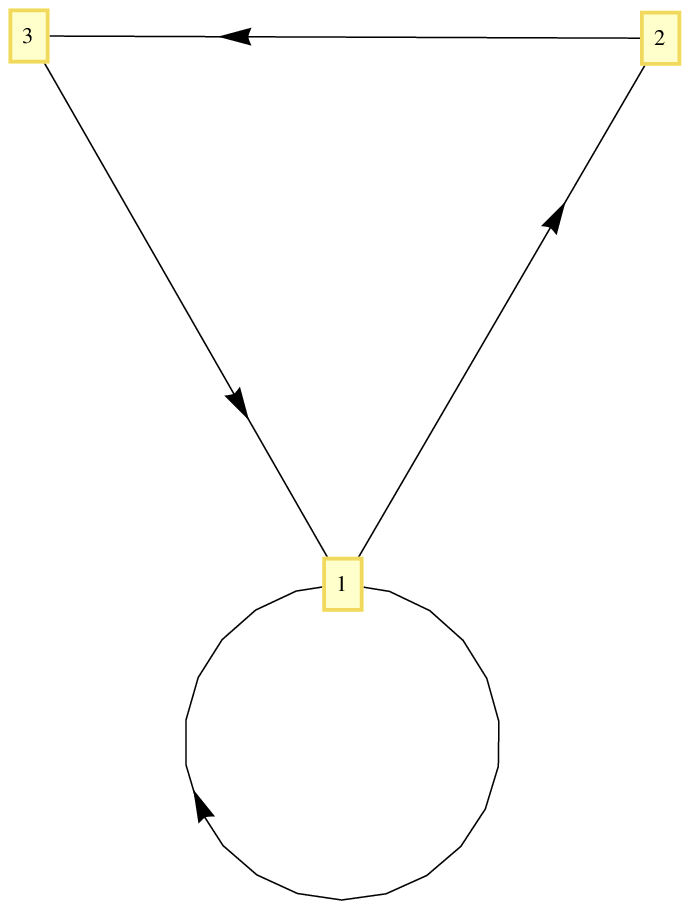}}
	& NO
	& \begin{tabular}{c|c} 0 & 4\end{tabular}
	& \begin{tabular}{c|c} 1 & 4 \end{tabular}\\
\hline
\end{tabular}
\end{center}
\end{table}

\subsection{Three Node Quivers \label{s:compu-3n}}
We present data for quivers with $n=3$ in Tables~\ref{tb:3n-3e}, \ref{tb:3n-4e} and \ref{tb:3n-5e} using the same format as the previous section.   Up to $e=5$ there are few enough theories to obtain an exact catalog of the VMS spaces.  For clarity we include all unique quivers, although as discussed in section \ref{s:alg-eg}, quivers with interchangeable adjoint fields have the same content.

\begin{table}[!]
\caption{\sf VMS Statistics for $(3,5)$ Quivers in Basic Case \label{tb:3n-5e}}
\begin{center}
\begin{tabular}{|c|c|c|c|}
\hline
	QUIVER & SAMPLING & Dim $|$ Count & Deg $|$ Count\\
\hline
	\raisebox{-0.4\height}{\includegraphics[width=4cm]{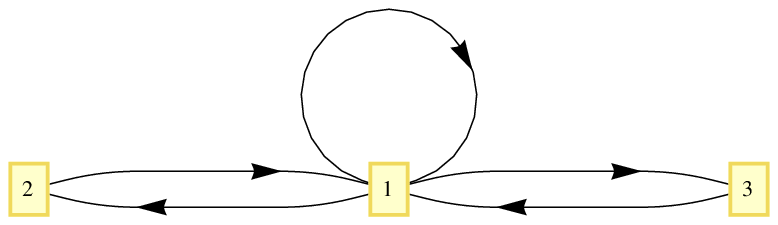}}
	& NO
	& \begin{tabular}{c|c} 1 & 2\\ 2 & 2 \end{tabular}
	& \begin{tabular}{c|c} 1 & 4 \end{tabular}\\
\hline
	\raisebox{-0.4\height}{\includegraphics[width=4cm]{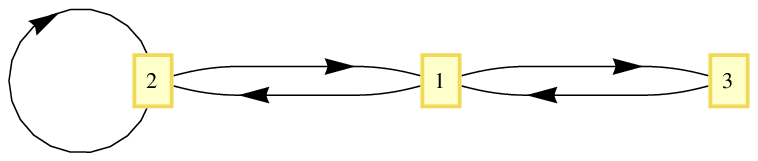}}
	& NO
	& \begin{tabular}{c|c} 1 & 1 \end{tabular}
	& \begin{tabular}{c|c} 1 & 1 \end{tabular}\\
\hline
\raisebox{-0.4\height}{\includegraphics[width=4cm]{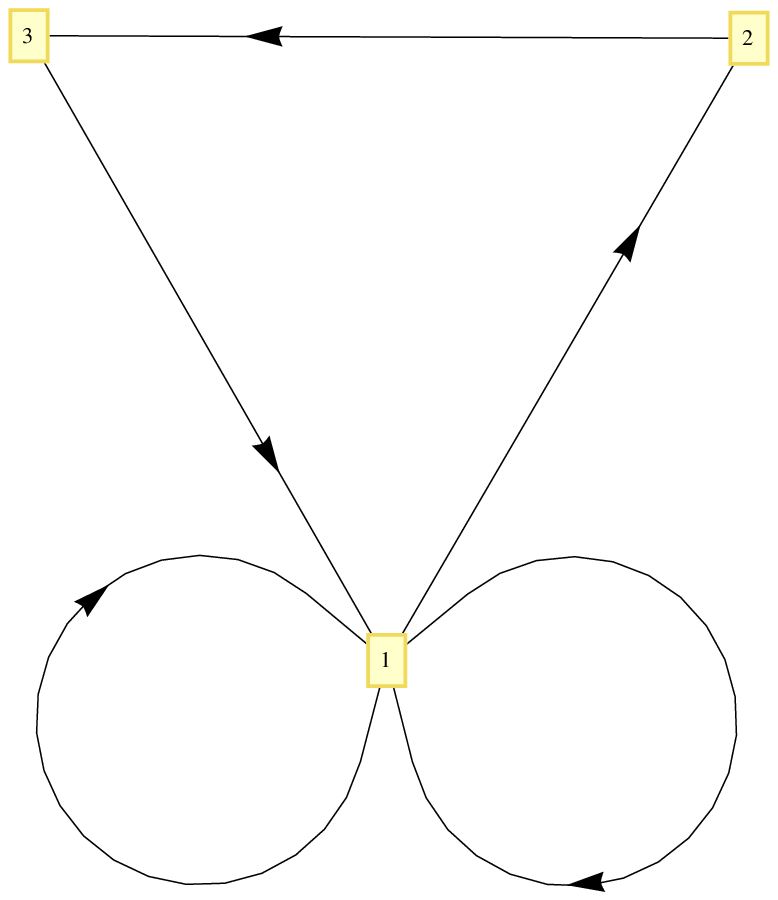}}
	& NO
	& \begin{tabular}{c|c} 0 & 31\\ 1 & 9 \end{tabular}
	& \begin{tabular}{c|c} 1 & 40 \end{tabular}\\
\hline
\raisebox{-0.4\height}{\includegraphics[width=4cm]{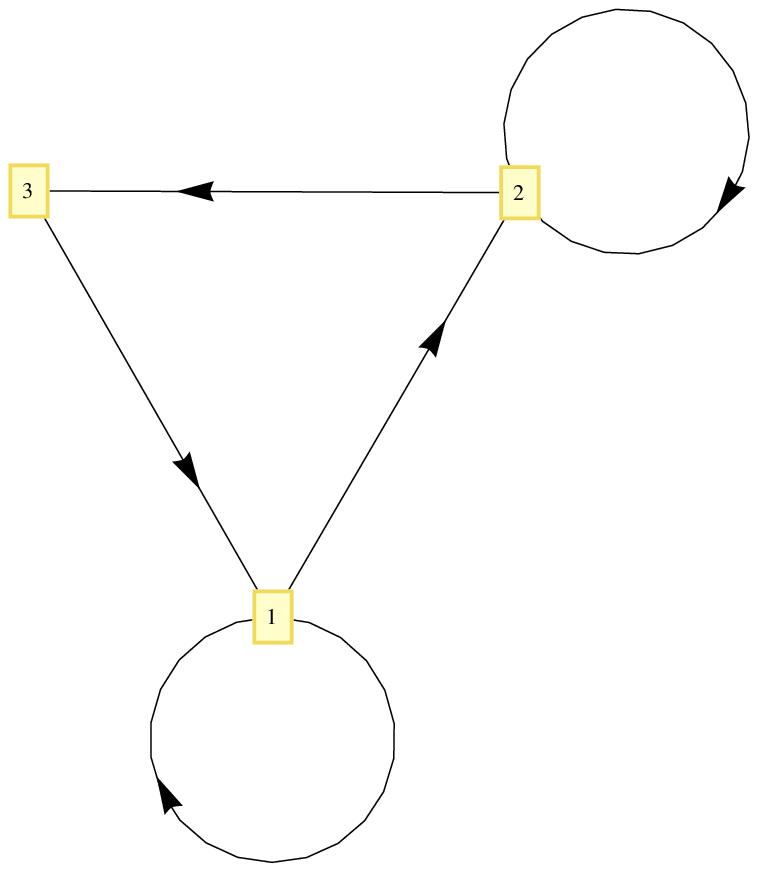}}
	& NO
	& \begin{tabular}{c|c} 0 & 31\\ 1 & 9 \end{tabular}
	& \begin{tabular}{c|c} 1 & 40 \end{tabular}\\
\hline
\raisebox{-0.4\height}{\includegraphics[width=4cm]{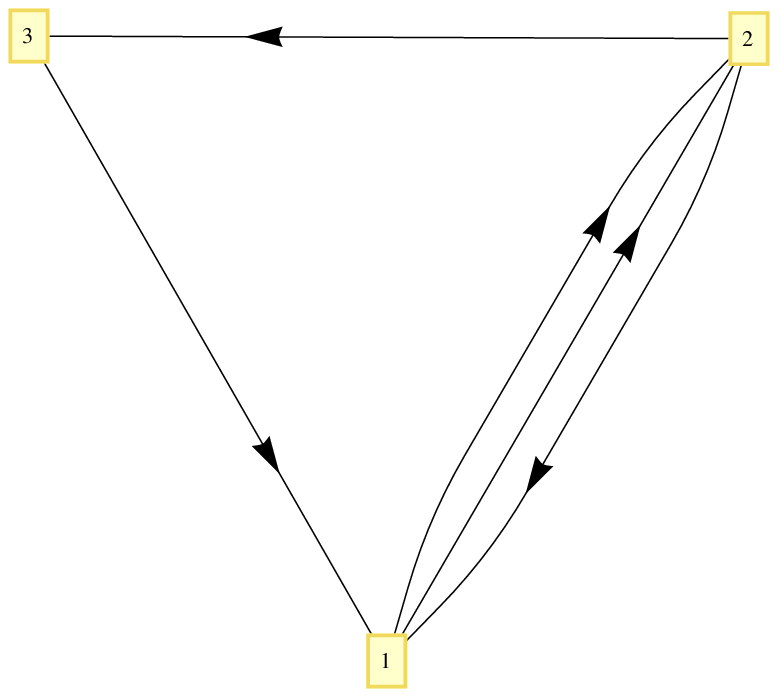}}
	& NO
	& \begin{tabular}{c|c} 2 & 4\end{tabular}
	& \begin{tabular}{c|c} 1 & 4 \end{tabular}\\
\hline
\end{tabular}
\end{center}
\end{table}

\subsection{Larger Data Sets \label{s:compu-lds}}
Of particular interest are the statistics regarding the number of
Calabi-Yau manifolds of complex dimension three for various diagrams.
We require large data sets to gain a clearer picture of the distribution
of these spaces among the vacuum moduli of the quiver theories.

Presented in Tables~\ref{tb:NGDIM}, \ref{tb:NGDEG}, \ref{tb:GDIM} and \ref{tb:GDEG} is some further data generated in order to initiate the construction of such a picture. Tables~\ref{tb:NGDIM} and \ref{tb:NGDEG} list the data for the basic cases, i.e. the coefficient of each GIO term can only take a value in $\{-1,0,1\}$. The remaining two tables list data for the generic case where the coefficients take integral values from $-5$ to $5$.  The sample size is chosen to be $s = 500$ as before. For each pair $(n, e)$ we list the counts of spaces of particular degree or dimension respectively.

In order to present this information, we combined the data generated for all  quivers with a given $(n,e)$.  In cases where random samples were taken for some or all the quivers, we needed to estimate an actual frequency of occurrence in the population of theories for $(n,e)$.  For example, if the quiver $Q$ 
generates $t_Q$ superpotentials, and we find $m_Q(d)$ superpotentials with dimension $d$, the 
proportion of theories with dimension $d$ is estimated via
\begin{equation}\label{eq:prop}
P(d) = \frac{\displaystyle \sum_Q \frac{m_Q(d)}{s} \cdot t_Q}{\displaystyle \sum_Q t_Q}\,.
\end{equation}

\begin{center}
\LTcapwidth=\textwidth
\begin{longtable}{|c|c|c|c|c|c|}
\caption{\sf Dimension of VMS for Pairs $(n, e)$ in the Basic Case: Given a pair $(n, e)$ we present tables containing all possible values of dim($\cal M$) in the left column and the corresponding proportion of theories for each dimension on the right. we set $s=200$ where random samples were required for quivers with $n=10,\ e=13$. \label{tb:NGDIM}} \\

\hline
\multicolumn{1}{|c|}{$n$} & \multicolumn{1}{c|}{$e=n$}
& \multicolumn{1}{c|}{$e=n+1$} & \multicolumn{1}{c|}{$e=n+2$}
& \multicolumn{1}{c|}{$e=n+3$} & \multicolumn{1}{c|}{$e=n+4$} \\
\endfirsthead

\hline
\multicolumn{1}{|c|}{$n$} & \multicolumn{1}{c|}{$e=n$}
& \multicolumn{1}{c|}{$e=n+1$} & \multicolumn{1}{c|}{$e=n+2$}
& \multicolumn{1}{c|}{$e=n+3$} & \multicolumn{1}{c|}{$e=n+4$} \\
\endhead

\multicolumn{6}{|r|}{\it\footnotesize{Continued on next page $\rightarrow$}} \\
\hline
\endfoot

\hline
\endlastfoot

\hline
2	& N/A
	& \begin{tabular}{c|c} 0 & 100\% \end{tabular}
	& \begin{tabular}{c|c} 0 & 61.5\%\\ 1 & 38.5\% \end{tabular}
	& \begin{tabular}{c|c} 0 & 76.8\%\\ 1 & 8.4\%\\ 2 & 14.8\% \end{tabular}
	& \begin{tabular}{c|c} 0 & 69.9\%\\ 1 & 28.2\% \\ 2 & 1.5\% \\ 3 & 0.4\% \end{tabular}\\
\hline
3	& \begin{tabular}{c|c} 0 & 100\% \end{tabular}
	& \begin{tabular}{c|c} 0 & 100\% \end{tabular}
	& \begin{tabular}{c|c} 0 & 69.7\%\\ 1 & 23.6\%\\ 2 & 6.7\% \end{tabular}
	& \begin{tabular}{c|c} 0 & 80.8\%\\ 1 & 9.3\%\\ 2 & 9.3\%\\ 3 & 0.6\% \end{tabular}
	& \begin{tabular}{c|c} 0 & 72.7\%\\ 1 & 24.2\%\\ 2 & 2.8\%\\ 3 & 0.2\%\\ 4 & 0.003\% \end{tabular}\\
\hline
4	& \begin{tabular}{c|c} 0 & 100\% \end{tabular}
	& \begin{tabular}{c|c} 0 & 80\%\\ 1 & 20\% \end{tabular}
	& \begin{tabular}{c|c} 0 & 60.9\%\\ 1 & 25.7\%\\ 2 & 13.4\% \end{tabular}
	& \begin{tabular}{c|c} 0 & 67.1\%\\ 1 & 22.0\%\\ 2 & 9.4\%\\ 3 & 1.5\% \end{tabular}
	& \begin{tabular}{c|c} 0 & 65.8\%\\ 1 & 27.1\%\\ 2 & 6.8\%\\ 3 & 0.4\%\\ 4 & 0.03\% \end{tabular}\\
\hline
5	& \begin{tabular}{c|c} 0 & 100\% \end{tabular}
	& \begin{tabular}{c|c} 0 & 66.7\%\\ 1 & 33.3\% \end{tabular}
	& \begin{tabular}{c|c} 0 & 56.8\%\\ 1 & 27.9\%\\ 2 & 15.2\% \end{tabular}
	& \begin{tabular}{c|c} 0 & 58.4\%\\ 1 & 27.2\%\\ 2 & 12.4\%\\ 3 & 2.0\% \end{tabular}
	& \begin{tabular}{c|c} 0 & 59.4\%\\ 1 & 30.1\%\\ 2 & 9.9\%\\ 3 & 0.6\%\\ 4 & 0.01\% \end{tabular}\\
\hline
6	& \begin{tabular}{c|c} 0 & 100\% \end{tabular}
	& \begin{tabular}{c|c} 0 & 66.7\%\\ 1 & 33.3\% \end{tabular}
	& \begin{tabular}{c|c} 0 & 50.0\%\\ 1 & 32.1\%\\ 2 & 17.9\% \end{tabular}
	& \begin{tabular}{c|c} 0 & 51.6\%\\ 1 & 30.4\%\\ 2 & 15.5\%\\ 3 & 2.6\% \end{tabular}
	& \begin{tabular}{c|c} 0 & 52.8\%\\ 1 & 33.3\%\\ 2 & 12.7\%\\ 3 & 1.2\%\\ 4 & 0.008\% \end{tabular}\\
\hline
7	& \begin{tabular}{c|c} 0 & 100\% \end{tabular}
	& \begin{tabular}{c|c} 0 & 61.5\%\\ 1 & 38.5\% \end{tabular}
	& \begin{tabular}{c|c} 0 & 47.8\%\\ 1 & 33.7\%\\ 2 & 18.5\% \end{tabular}
	& \begin{tabular}{c|c} 0 & 47.2\%\\ 1 & 32.1\%\\ 2 & 17.7\%\\ 3 & 3.0\% \end{tabular}
	& N/A \\
\hline
8	& \begin{tabular}{c|c} 0 & 100\% \end{tabular}
	& \begin{tabular}{c|c} 0 & 61.5\%\\ 1 & 38.5\% \end{tabular}
	& \begin{tabular}{c|c} 0 & 43.8\%\\ 1 & 36.4\%\\ 2 & 19.8\% \end{tabular}
	& \begin{tabular}{c|c} 0 & 42.9\%\\ 1 & 33.9\%\\ 2 & 19.9\%\\ 3 & 3.4\% \end{tabular}
	& N/A \\
\hline
9	& \begin{tabular}{c|c} 0 & 100\% \end{tabular}
	& \begin{tabular}{c|c} 0 & 58.8\%\\ 1 & 41.2\% \end{tabular}
	& \begin{tabular}{c|c} 0 & 42.0\%\\ 1 & 37.7\%\\ 2 & 20.3\% \end{tabular}
	& \begin{tabular}{c|c} 0 & 38.5\%\\ 1 & 35.5\%\\ 2 & 22.1\%\\ 3 & 3.9\% \end{tabular}
	& N/A \\
\hline
10	& \begin{tabular}{c|c} 0 & 100\% \end{tabular}
	& \begin{tabular}{c|c} 0 & 58.8\%\\ 1 & 41.2\% \end{tabular}
	& \begin{tabular}{c|c} 0 & 39.2\%\\ 1 & 39.7\%\\ 2 & 21.1\% \end{tabular}
	& \begin{tabular}{c|c} 0 & 35.7\%\\ 1 & 36.1\%\\ 2 & 23.9\%\\ 3 & 4.3\% \end{tabular}
	& N/A \\
\end{longtable}
\end{center}

As expected, we observe more complex vacuum structure as we increase the number of edges for a given quiver with $n$ nodes in Table~\ref{tb:NGDIM}. Fixing $(e-n)$, the data reveals some interesting trends with increasing $n$. The proportion of zero dimensional spaces decreases significantly from roughly 80\% to 40\% for $e-n=3$ for example. And the largest dimension is congruously $(e-n)$. Increasing the number of nodes gives us more theories with non-vanishing dimension and richer structure. 

In Table~\ref{tb:NGDEG}, the distribution of the VMS degree on the whole exhibits similar features to the dimension distribution.  In our set of quivers we found theories with degree up to 10, although higher degree vacuums may exist that did not appear in our samples.

\begin{center}
\LTcapwidth=\textwidth
\begin{longtable}{|c|c|c|c|c|c|}
\caption{\sf Degree of VMS for pairs $(n,e)$ in the Basic Case: Given a pair $(n, e)$ we present tables containing all possible values of deg($\cal M$) in the left column and the corresponding proportion of theories for each degree on the right. We set $s=200$ where random samples were required for quivers with $n=10,\ e=13$. \label{tb:NGDEG}}\\

\hline
\multicolumn{1}{|c|}{$n$} & \multicolumn{1}{c|}{$e=n$}
& \multicolumn{1}{c|}{$e=n+1$} & \multicolumn{1}{c|}{$e=n+2$}
& \multicolumn{1}{c|}{$e=n+3$} & \multicolumn{1}{c|}{$e=n+4$} \\
\endfirsthead

\hline
\multicolumn{1}{|c|}{$n$} & \multicolumn{1}{c|}{$e=n$}
& \multicolumn{1}{c|}{$e=n+1$} & \multicolumn{1}{c|}{$e=n+2$}
& \multicolumn{1}{c|}{$e=n+3$} & \multicolumn{1}{c|}{$e=n+4$} \\
\endhead

\multicolumn{6}{|r|}{\it\footnotesize{Continued on next page $\rightarrow$}} \\
\hline
\endfoot

\hline
\endlastfoot

\hline
2	& N/A
	& \begin{tabular}{c|c} 1 & 100\% \end{tabular}
	& \begin{tabular}{c|c} 1 & 100\% \end{tabular}
	& \begin{tabular}{c|c} 1 & 86.5\%\\ 2 & 10.8\%\\ 3 & 2.7\% \end{tabular}
	& \begin{tabular}{c|c} 1 & 81.0\%\\ 2 & 14.6\%\\ 3 & 2.9\%\\ 4 & 1.0\%\\ 5 & 0.1\%\\
	6 & 0.3\%\\ 10 & 0.07\%
	\end{tabular}\\
\hline
3	& \begin{tabular}{c|c} 1 & 100\% \end{tabular}
	& \begin{tabular}{c|c} 1 & 100\% \end{tabular}
	& \begin{tabular}{c|c} 1 & 100\% \end{tabular}
	& \begin{tabular}{c|c} 1 & 91.0\%\\ 2 & 7.9\%\\ 3 & 1.1\% \end{tabular}
	& \begin{tabular}{c|c} 1 & 82.0\%\\ 2 & 14.0\%\\ 3 & 2.7\%\\ 4 & 1.0\%\\ 5 & 0.1\%\\
	6 & 0.2\%\\ 7 & 0.0001\%\\ 8 & 0.01\% \end{tabular}\\
\hline
4	& \begin{tabular}{c|c} 1 & 100\% \end{tabular}
	& \begin{tabular}{c|c} 1 & 100\% \end{tabular}
	& \begin{tabular}{c|c} 1 & 98.5\%\\ 2 & 1.5\% \end{tabular}
	& \begin{tabular}{c|c} 1 & 90.1\%\\ 2 & 9.0\%\\ 3 & 0.8\%\\ 4 & 0.02\% \end{tabular}
	& \begin{tabular}{c|c} 1 & 77.1\%\\ 2 & 17.6\%\\ 3 & 3.8\%\\ 4 & 1.1\%\\ 5 & 0.2\%\\
	6 & 0.2\%\\ 7 & 0.0002\%\\ 8 & 0.000004\%\\ 10 & 0.03\% \end{tabular}\\
\hline
5	& \begin{tabular}{c|c} 1 & 100\% \end{tabular}
	& \begin{tabular}{c|c} 1 & 100\% \end{tabular}
	& \begin{tabular}{c|c} 1 & 99.0\%\\ 2 & 1.0\% \end{tabular}
	& \begin{tabular}{c|c} 1 & 88.3\%\\ 2 & 10.4\%\\ 3 & 1.2\%\\ 4 & 0.1\% \end{tabular}
	& \begin{tabular}{c|c} 1 & 74.3\%\\ 2 & 19.8\%\\ 3 & 4.2\%\\ 4 & 1.3\%\\ 5 & 0.2\%\\
	6 & 0.2\%\\ 7 & 0.0001\%\\ 8 & 0.008\%\\ 10 & 0.01\% \end{tabular}\\
\hline
6	& \begin{tabular}{c|c} 1 & 100\% \end{tabular}
	& \begin{tabular}{c|c} 1 & 100\% \end{tabular}
	& \begin{tabular}{c|c} 1 & 97.4\%\\ 2 & 2.6\% \end{tabular}
	& \begin{tabular}{c|c} 1 & 86.4\%\\ 2 & 12.3\%\\ 3 & 1.2\%\\ 4 & 0.1\% \end{tabular}
	& \begin{tabular}{c|c} 1 & 71.1\%\\ 2 & 21.5\%\\ 3 & 5.0\%\\ 4 & 1.7\%\\ 5 & 0.3\%\\
		6 & 0.2\%\\ 7 & 0.01\%\\ 8 & 0.01\%\\ 10 & 0.007\% \end{tabular}\\
\hline
7	& \begin{tabular}{c|c} 1 & 100\% \end{tabular}
	& \begin{tabular}{c|c} 1 & 100\% \end{tabular}
	& \begin{tabular}{c|c} 1 & 97.6\%\\ 2 & 2.4\% \end{tabular}
	& \begin{tabular}{c|c} 1 & 85.6\%\\ 2 & 12.9\%\\ 3 & 1.3\%\\ 4 & 0.2\% \end{tabular}
	& N/A\\
\hline
8	& \begin{tabular}{c|c} 1 & 100\% \end{tabular}
	& \begin{tabular}{c|c} 1 & 100\% \end{tabular}
	& \begin{tabular}{c|c} 1 & 96.6\%\\ 2 & 3.4\% \end{tabular}
	& \begin{tabular}{c|c} 1 & 84.0\%\\ 2 & 14.1\%\\ 3 & 1.6\%\\ 4 & 0.2\% \end{tabular}
	& N/A\\
\hline
9	& \begin{tabular}{c|c} 1 & 100\% \end{tabular}
	& \begin{tabular}{c|c} 1 & 100\% \end{tabular}
	& \begin{tabular}{c|c} 1 & 96.5\%\\ 2 & 3.5\% \end{tabular}
	& \begin{tabular}{c|c} 1 & 82.2\%\\ 2 & 15.4\%\\ 3 & 2.2\%\\ 4 & 0.3\%\\ 6 & 0.01\% \end{tabular}
	& N/A\\
\hline
10	& \begin{tabular}{c|c} 1 & 100\% \end{tabular}
	& \begin{tabular}{c|c} 1 & 100\% \end{tabular}
	& \begin{tabular}{c|c} 1 & 95.7\%\\ 2 & 4.3\% \end{tabular}
	& \begin{tabular}{c|c} 1 & 80.9\%\\ 2 & 16.5\%\\ 3 & 2.3\%\\ 4 & 0.3\%\\ 6 & 0.02 \end{tabular}
	& N/A\\
\end{longtable}
\end{center}

Expanding our set of possible coefficients for the generic case resulted in a more limited range of spaces in our results shown in Tables~\ref{tb:GDIM} and \ref{tb:GDEG}.  Although the number of theories increased for each quiver, our sample no longer included spaces with the higher dimension or degree values seen in the basic case.   For example spaces of dimension 4 and degree above 3 that occurred in samples for unit coefficients were no longer present. 

The proportion of spaces with dimension zero is notably higher in these tables compared to the basic case.  We infer that the distributions are skewed further towards lower degree and dimension when generic coefficients are allowed.

\begin{center}
\LTcapwidth=\textwidth
\begin{longtable}{|c|c|c|c|c|}
\caption{{\sf Dimension of VMS for Pairs $(n,e)$ in Generic Case \label{tb:GDIM}}} \\

\hline
\multicolumn{1}{|c|}{$n$} & \multicolumn{1}{c|}{$e=n$}
& \multicolumn{1}{c|}{$e=n+1$} & \multicolumn{1}{c|}{$e=n+2$}
& \multicolumn{1}{c|}{$e=n+3$} \\
\endfirsthead

\hline
\multicolumn{1}{|c|}{$n$} & \multicolumn{1}{c|}{$e=n$}
& \multicolumn{1}{c|}{$e=n+1$} & \multicolumn{1}{c|}{$e=n+2$}
& \multicolumn{1}{c|}{$e=n+3$} \\
\endhead

\multicolumn{5}{|r|}{\it\footnotesize{Continued on next page $\rightarrow$}} \\
\hline
\endfoot

\hline
\endlastfoot

\hline
2	& N/A
	& \begin{tabular}{c|c} 0 & 100\% \end{tabular}
	& \begin{tabular}{c|c} 0 & 84.7\%\\ 1 & 15.3\% \end{tabular}
	& \begin{tabular}{c|c} 0 & 98.3\%\\ 1 & 1.4\%\\ 2 & 0.3\% \end{tabular}\\
\hline
3	& \begin{tabular}{c|c} 0 & 100\% \end{tabular}
	& \begin{tabular}{c|c} 0 & 100\% \end{tabular}
	& \begin{tabular}{c|c} 0 & 95.5\%\\ 1 & 4.0\%\\ 2 & 0.5\% \end{tabular}
	& \begin{tabular}{c|c} 0 & 98.5\%\\ 1 & 1.4\%\\ 2 & 0.1\%\\ 3 & 0.001\% \end{tabular}\\
\hline
4	& \begin{tabular}{c|c} 0 & 100\% \end{tabular}
	& \begin{tabular}{c|c} 0 & 92.3\%\\ 1 & 7.7\% \end{tabular}
	& \begin{tabular}{c|c} 0 & 92.7\%\\ 1 & 6.2\%\\ 2 & 1.1\% \end{tabular}
	& \begin{tabular}{c|c} 0 & 95.3\%\\ 1 & 4.6\%\\ 2 & 0.1\%\\ 3 & 0.002\% \end{tabular}\\
\hline
5	& \begin{tabular}{c|c} 0 & 100\% \end{tabular}
	& \begin{tabular}{c|c} 0 & 88\%\\ 1 & 12\% \end{tabular}
	& \begin{tabular}{c|c} 0 & 90.9\%\\ 1 & 7.7\%\\ 2 & 1.4\% \end{tabular}
	& \begin{tabular}{c|c} 0 & 93.0\%\\ 1 & 6.6\%\\ 2 & 0.4\%\\ 3 & 0.02\% \end{tabular}\\
\hline
6	& \begin{tabular}{c|c} 0 & 100\% \end{tabular}
	& \begin{tabular}{c|c} 0 & 88\%\\ 1 & 12\% \end{tabular}
	& \begin{tabular}{c|c} 0 & 86.7\%\\ 1 & 11.4\%\\ 2 & 1.9\% \end{tabular}
	& \begin{tabular}{c|c} 0 & 91\%\\ 1 & 8\%\\ 2 & 0.9\%\\ 3 & 0.05\% \end{tabular}\\
\hline
7	& \begin{tabular}{c|c} 0 & 100\% \end{tabular}
	& \begin{tabular}{c|c} 0 & 86.5\%\\ 1 & 13.5\% \end{tabular}
	& \begin{tabular}{c|c} 0 & 87.0\%\\ 1 & 11.3\%\\ 2 & 1.7\% \end{tabular}
	& N/A\\
\hline
8	& \begin{tabular}{c|c} 0 & 100\% \end{tabular}
	& \begin{tabular}{c|c} 0 & 86.5\%\\ 1 & 13.5\% \end{tabular}
	& \begin{tabular}{c|c} 0 & 84.2\%\\ 1 & 13.9\%\\ 2 & 1.9\% \end{tabular}
	& N/A\\
\hline
9	& \begin{tabular}{c|c} 0 & 100\% \end{tabular}
	& \begin{tabular}{c|c} 0 & 85.7\%\\ 1 & 14.3\% \end{tabular}
	& \begin{tabular}{c|c} 0 & 82.4\%\\ 1 & 15.5\%\\ 2 & 2.1\% \end{tabular}
	& N/A\\
\hline
10	& \begin{tabular}{c|c} 0 & 100\% \end{tabular}
	& \begin{tabular}{c|c} 0 & 85.7\%\\ 1 & 14.3\% \end{tabular}
	& \begin{tabular}{c|c} 0 & 81.2\%\\ 1 & 16.8\%\\ 2 & 2.1\% \end{tabular}
	& N/A\\
\hline
\end{longtable}
\end{center}

\begin{center}
\LTcapwidth=\textwidth
\begin{longtable}{|c|c|c|c|c|}
\caption{\sf Degree of VMS for Pairs $(n,e)$ in Generic Case \label{tb:GDEG}} \\

\hline
\multicolumn{1}{|c|}{$n$} & \multicolumn{1}{c|}{$e=n$}
& \multicolumn{1}{c|}{$e=n+1$} & \multicolumn{1}{c|}{$e=n+2$}
& \multicolumn{1}{c|}{$e=n+3$} \\
\endfirsthead

\hline
\multicolumn{1}{|c|}{$n$} & \multicolumn{1}{c|}{$e=n$}
& \multicolumn{1}{c|}{$e=n+1$} & \multicolumn{1}{c|}{$e=n+2$}
& \multicolumn{1}{c|}{$e=n+3$} \\
\endhead

\multicolumn{5}{|r|}{\it\footnotesize{Continued on next page $\rightarrow$}} \\
\hline
\endfoot

\hline
\endlastfoot

\hline
2	& N/A
	& \begin{tabular}{c|c} 1 & 100\% \end{tabular}
	& \begin{tabular}{c|c} 1 & 100\% \end{tabular}
	& \begin{tabular}{c|c} 1 & 99.7\%\\ 2 & 0.3\%\\ 3 & 0.002\% \end{tabular}\\
\hline
3	& \begin{tabular}{c|c} 1 & 100\% \end{tabular}
	& \begin{tabular}{c|c} 1 & 100\% \end{tabular}
	& \begin{tabular}{c|c} 1 & 100\% \end{tabular}
	& \begin{tabular}{c|c} 1 & 99.9\%\\ 2 & 0.1\%\\ 3 & 0.003\% \end{tabular}\\
\hline
4	& \begin{tabular}{c|c} 1 & 100\% \end{tabular}
	& \begin{tabular}{c|c} 1 & 100\% \end{tabular}
	& \begin{tabular}{c|c} 1 & 99.97\%\\ 2 & 0.03\% \end{tabular}
	& \begin{tabular}{c|c} 1 & 99.9\%\\ 2 & 0.1\%\\ 3 & 0.0002\% \end{tabular}\\
\hline
5	& \begin{tabular}{c|c} 1 & 100\% \end{tabular}
	& \begin{tabular}{c|c} 1 & 100\% \end{tabular}
	& \begin{tabular}{c|c} 1 & 99.96\%\\ 2 & 0.04\% \end{tabular}
	& \begin{tabular}{c|c} 1 & 99.7\%\\ 2 & 0.3\%\\ 3 & 0.0001\% \end{tabular}\\
\hline
6	& \begin{tabular}{c|c} 1 & 100\% \end{tabular}
	& \begin{tabular}{c|c} 1 & 100\% \end{tabular}
	& \begin{tabular}{c|c} 1 & 99.8\%\\ 2 & 0.2\% \end{tabular}
	& \begin{tabular}{c|c} 1 & 99.6\%\\ 2 & 0.4\%\\ 3 & 0.0007\% \end{tabular}\\
\hline
7	& \begin{tabular}{c|c} 1 & 100\% \end{tabular}
	& \begin{tabular}{c|c} 1 & 100\% \end{tabular}
	& \begin{tabular}{c|c} 1 & 99.8\%\\ 2 & 0.2\% \end{tabular}
	& N/A\\
\hline
8	& \begin{tabular}{c|c} 1 & 100\% \end{tabular}
	& \begin{tabular}{c|c} 1 & 100\% \end{tabular}
	& \begin{tabular}{c|c} 1 & 99.6\%\\ 2 & 0.4\% \end{tabular}
	& N/A\\
\hline
9	& \begin{tabular}{c|c} 1 & 100\% \end{tabular}
	& \begin{tabular}{c|c} 1 & 100\% \end{tabular}
	& \begin{tabular}{c|c} 1 & 99.7\%\\ 2 & 0.3\% \end{tabular}
	& N/A\\
\hline
10	& \begin{tabular}{c|c} 1 & 100\% \end{tabular}
	& \begin{tabular}{c|c} 1 & 100\% \end{tabular}
	& \begin{tabular}{c|c} 1 & 99.6\%\\ 2 & 0.4\% \end{tabular}
	& N/A\\
\hline
\end{longtable}
\end{center}

\subsection{Degree-Dimension Scatter \label{s:compu-sca}}

Finally we collect our data from both the simple and generic cases respectively to demonstrate the relationship between degree and dimension. 

\begin{figure}[!t]
	\begin{tabular}{cc}
	\includegraphics[width=0.48\textwidth]{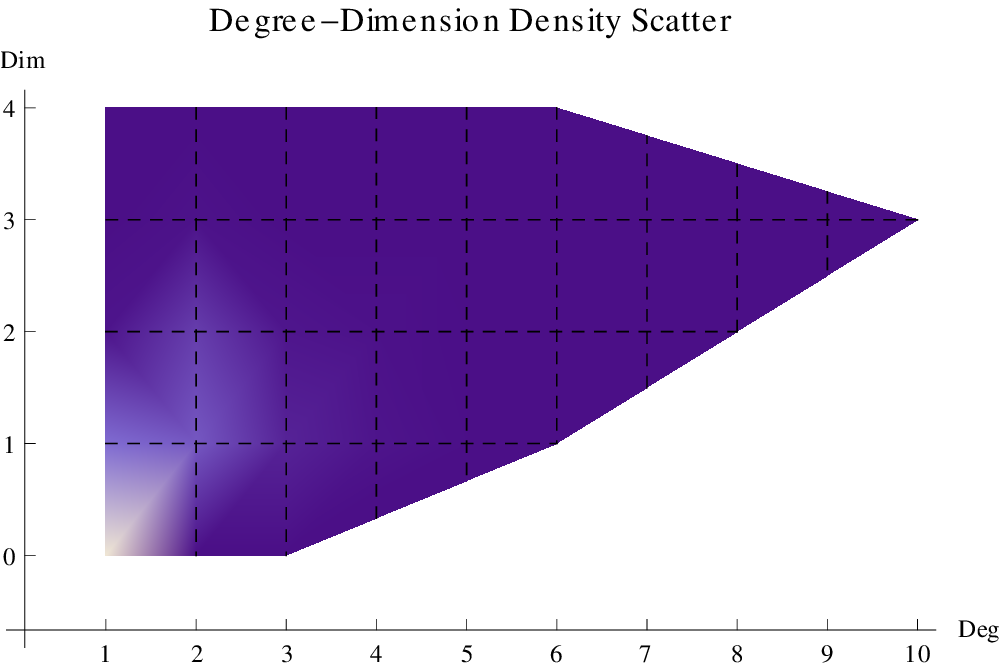}
	& \includegraphics[width=0.48\textwidth]{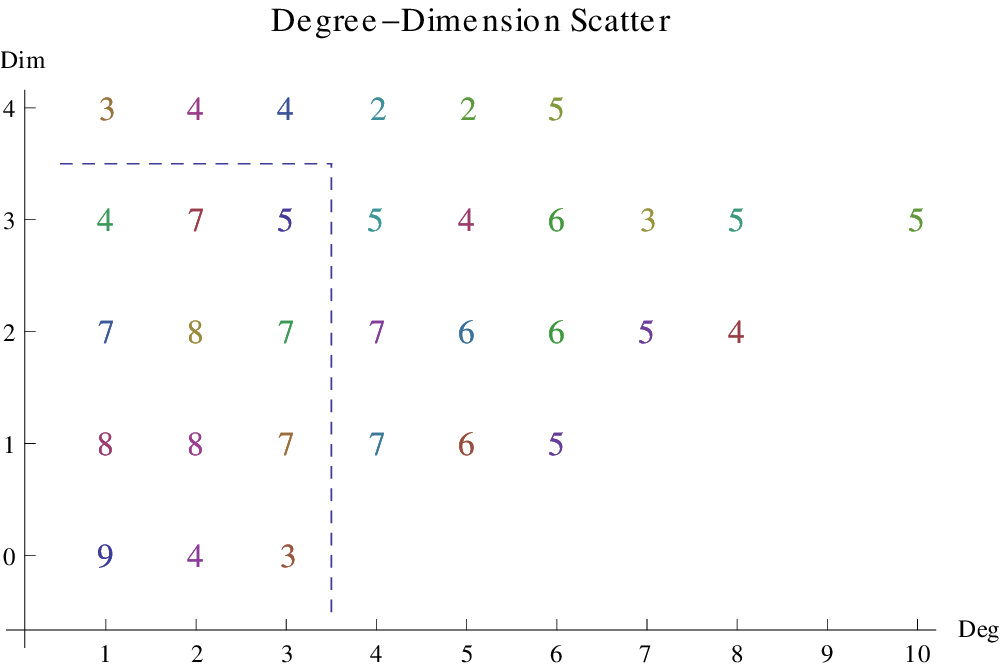}
	\end{tabular}
	\caption{{\sf Scatter Plot in Basic Case: On the left, dark to light indicates an increasing frequency of quiver theories for given pairs of dimension and degree. On the right, a figure $x$ indicates a raw frequency estimate for a given degree and dimension in the range $10^x$ to $10^{x+1}$.\label{f:non-gen-sca} }}
\end{figure}

\begin{figure}[!t]
	\begin{tabular}{cc}
	\includegraphics[width=0.48\textwidth]{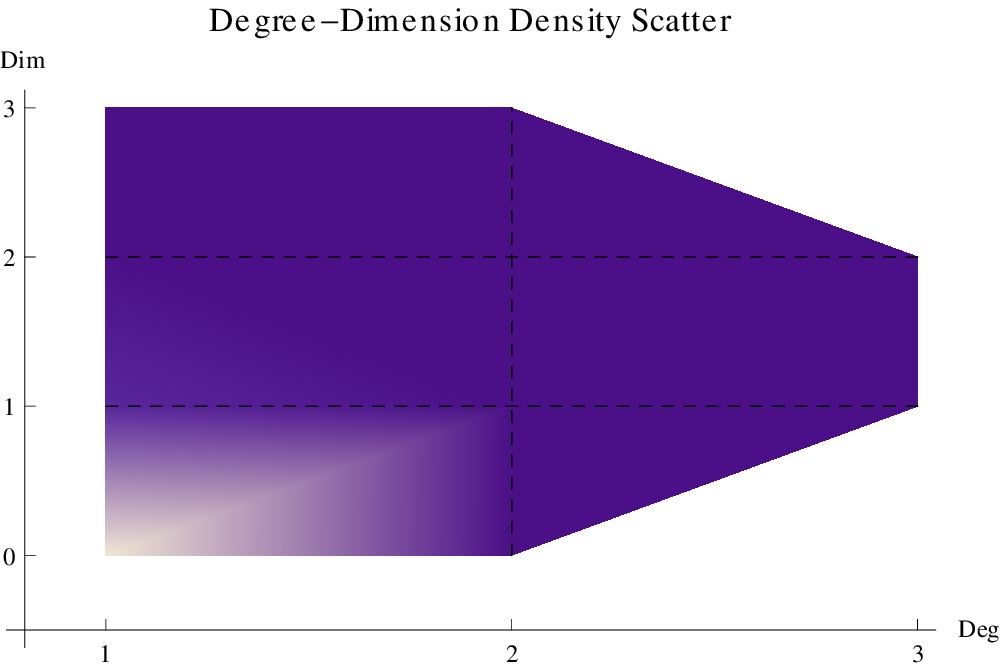}
	& \includegraphics[width=0.48\textwidth]{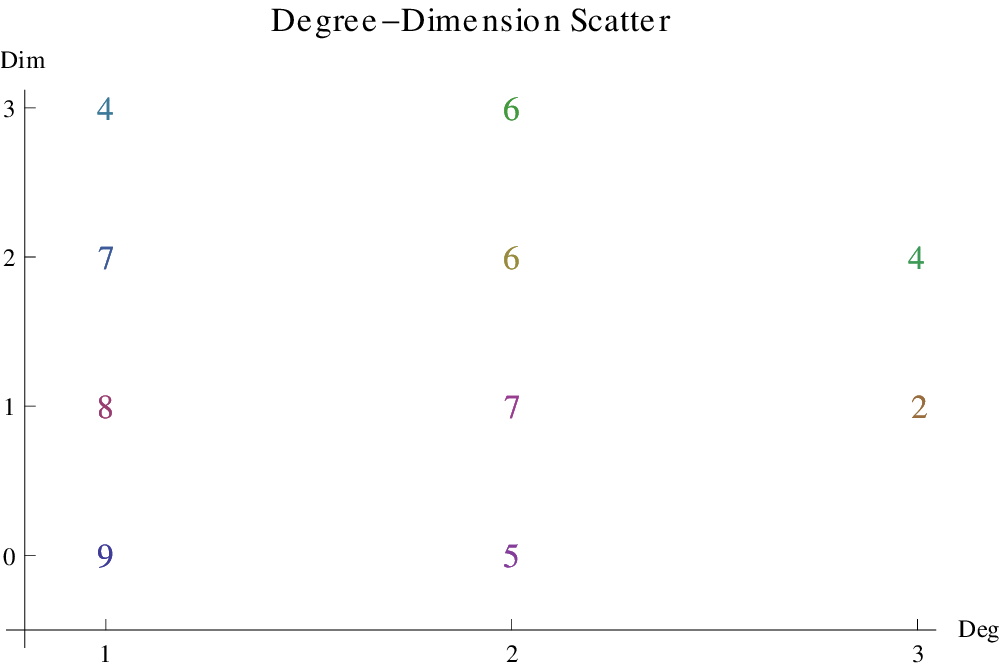}
	\end{tabular}
	\caption{{\sf Scatter Plot in Generic Case: On the left, dark to light indicates an increasing frequency of quiver theories for given pairs of dimension and degree. On the right, a figure $x$ indicates a raw frequency estimate for a given degree and dimension in the range $10^x$ to $10^{x+1}$. \label{f:gen-sca}}}
\end{figure}

The charts to the left in Figures~\ref{f:non-gen-sca} and \ref{f:gen-sca} give the density of the occurrence of various (deg, dim) pairs while we give estimates of the raw frequencies for each pair on the right.  Our estimates were calculated using
\begin{equation}
{\rm Freq}({\rm deg},{\rm dim}) = \sum_{n,\,e} \sum_Q \frac{m_Q({\rm deg},{\rm dim})}{s} \cdot t_Q\,,
\end{equation}
where $m_Q$, $s$ and $t_Q$ are the quantities previously defined in Eq.~(\ref{eq:prop}).

\subsection{Algebraic Geometry of VMS \label{s:compu-ana}}
It is expedient to step back and briefly see our VMS computation from the point of view of algebraic geometry.
For the completely generic situation of random complex coefficients, the dimension of the VMS can be readily estimated by counting the degrees of freedom (how many complex variables and how many constraining equations).
However, what we are doing here is a substantial refinement.
First, inspired by the ``toric condition'' for quiver theories, we have considered the restriction of coefficients to $\{0,\pm1\}$, what we refer to as the basic case, a sublocus of the generic case in the complex structure moduli space.

Moreover, we have also computed also the degree and, importantly, the Hilbert series of the variety, encoding the precise algebraic and birational information of the vacuum.
We recall that one of the most fundamental quantities which characterises an algebraic variety $X$ is the {\bf Hilbert series}, which is the generating function for the number $\dim(X_i)$ of independent polynomials at a given degree on $X$:
\begin{equation}\label{hilb}
H(t;~X) = \sum_{i=0}^\infty \dim(X_i) t^i = (1-t)^{-k} Q(t) =
(1-t)^{-\dim(X)} P(t) \ .
\end{equation}
In the above, $k$ is the number of variables in the defining equation of $X$ 
and most importantly, $P(t)$ and $Q(t)$ as defined are polynomials with integer 
coefficients.
The Hilbert series is key to the Plethystic programme of enumerating gauge invariant operators \cite{Benvenuti:2006qr}.

\begin{table}[!b]
\caption{\sf Percentages of Palindromic Hilbert Series for Pairs $(n,e)$ in Basic Case \label{tb:BPHS}}
\begin{center}
\begin{tabular}{|c|c|c|c|c|c|}
\hline
$n\backslash e$ & $e=n$ & $e=n+1$ & $e=n+2$ & $e=n+3$ & $e=n+4$ \\
\hline
2 & N/A & 100\%	& 100\%	& 96.5\% & 82.7\% \\
\hline
3 & 100\% & 100\% & 100\% &	97.4\% & 89.1\% \\
\hline
4 & 100\% & 100\% & 100\% & 98.1\% & 91.2\% \\
\hline
5 & 100\% & 100\% & 100\% & 97.9\% & 92.6\% \\
\hline
6 & 100\% & 100\% & 100\% & 97.6\% & 93.1\% \\
\hline
7 & 100\% & 100\% & 100\% & 97.3\% & N/A \\
\hline
8 & 100\% & 100\% & 100\% & 96.9\% & N/A \\
\hline
9 & 100\% & 100\% & 100\% & 96.3\% & N/A \\
\hline
10 & 100\% & 100\% & 100\% & 97.3\% & N/A \\
\hline
\end{tabular}
\end{center}
\end{table}

\begin{table}[!]
\caption{\sf Percentages of Palindromic Hilbert Series for Pairs $(n,e)$ in Generic Case \label{tb:GPHS}}
\begin{center}
\begin{tabular}{|c|c|c|c|c|}
\hline
$n\backslash e$ & $e=n$ & $e=n+1$ & $e=n+2$ & $e=n+3$ \\
\hline
2 & N/A & 100\%	& 100\%	& 98\% \\
\hline
3 & 100\% & 100\% & 100\% &	98.9\% \\
\hline
4 & 100\% & 100\% & 100\% & 99.7\% \\
\hline
5 & 100\% & 100\% & 100\% & 99.7\% \\
\hline
6 & 100\% & 100\% & 100\% & 99.8\% \\
\hline
7 & 100\% & 100\% & 100\% & N/A \\
\hline
8 & 100\% & 100\% & 100\% & N/A \\
\hline
9 & 100\% & 100\% & 100\% & N/A \\
\hline
10 & 100\% & 100\% & 100\% & N/A \\
\hline
\end{tabular}
\end{center}
\end{table}

We generated the Hilbert series using
Macaulay2 for all the quivers investigated in this paper for a more detailed catalog of VMS spaces.
The results are too extensive to be included within this document. Instead we have provided the full information at
\href{https://github.com/dayzhou/Hilbert-Series}{GitHub}\footnote{\tt https://github.com/dayzhou/Hilbert-Series}
by listing all possible Hilbert series and
their prevalence for each $(n, e)$. Both the basic and generic cases are available in both a readable format and as raw data.

In tables \ref{tb:BPHS} and \ref{tb:GPHS}, we present the statistical data of palindromic Hilber series in percentages for both basic and generic cases. Now, importantly, when the numerator of the Hilbert series is palindromic, i.e., the coefficients $p_i$ are such that $p_i = p_{n-i}$ for all $i$, a theorem of Stanley \cite{stanley} guarantees that the algebraic variety is Calabi-Yau; this was first used in the gauge theory context in \cite{Forcella:2008bb}.
Scanning over our database of 926883 Hilbert series in basic case and 270410 in generic case, we find that 879852 and 270052 in each case have this property. We conclude that
{\it 94.9\% of our VMS are Calabi-Yau in basic case, and 99.9\% in generic case}.
It is interesting to see that with increasing complexity of the theory (increasing number of fields and gauge groups), the chances of it being Calabi-Yau seems to decrease. We can also see from our statistics that restricting the coefficients in superpotentials from $\{-5,\cdots,5\}$ to $\{-1,0,1\}$ reduces the probability of being Calabi-Yau by about 5\%.

\section{Conclusions and Prospects \label{s:conc}}
Given the vast landscape of quantum field theories, especially those with ${\cal N}=1$ supersymmetry whose matter content and interaction abound untameably, and given their underlying geometry in the form of the vacuum moduli space whose efficacy has ranged from phenomenology to holography, it is clearly desirable to investigate the space of such vacua.
In this paper, we have initiated the study of the statistics of this space of vacuum moduli spaces, in the distribution of the dimension and degree as algebraic varieties and have explicitly computed the Hilbert series for thousands of samples.

We find that the representative vacuum for our study of $n\leq 10$ and $e\leq n+4$ quivers has low dimension and degree: our distributions appear highly skewed particularly for the generic case.  Although increasing $n$ produces more interesting phenomena, the spaces of complex dimension three remain $\leq5\%$ of the quivers for $(n,e)$ in our study, validating our approach as heavily selective of the vacuum landscape.  Our catalog of Hilbert series elucidate deeper information: the chances of the VMS being Calabi-Yau decreases as $(e-n)$ increases or when we we restrict the coefficient set to $\{-1,0,1\}$.

Indeed, we have only begun a clearly profitable enterprise.
Throughout this paper we have used Abelian quiver gauge theories (in the basic (toric) case and the generic case) as a testing ground, one can extend this to more general quivers and to field theories admitting more than bifundamental and adjoints. Physically, the class of such theories is by far the most studied, especially in the context of string theory and AdS/CFT because of the underlying toric Calabi-Yau geometry.
Mathematically, it is interesting to point out that {\it every} projective variety (note that all our affine varieties are trivially complex cones over some projective variety) is a quiver Grassmannian \cite{savage}.
Thus, from both physical and mathematical motivations, our representative class of theories is highly non-trivial.

Recently, there has been nice papers \cite{Greene:2013ida, Aravind:2014} studying the instabilities in the landscape of high-dimensional moduli spaces.
Specifically, as stated in \cite{Greene:2013ida}, ``tunneling rates, and hence vacuum instability, grow so rapidly with the number of moduli that the probability of a given local minimum being metastable is exponentially small".
Our statistical outlook is very much in this spirit. It would be interesting to extract such physical properties from our data.
For example, our data shows that the proportion of large dimensional moduli spaces (the real/global vacua of the theories) in all moduli spaces is considerably small, which is consistent with the conclusion of \cite{Greene:2013ida}.
Indeed, a programme of using vacuum geometry for the sake of particle phenomenology of the standard model and beyond has been ongoing \cite{Gray:2005sr,Gray:2006jb,Hanany:2010vu,cyril}. Our geometric data should be useful toward the question of how {\it geometrically generic} the (supersymmetric) standard model is.

\newpage

\section*{Acknowledgements}
We would like
to thank Zhi-Guang Xiao and the Cloud Lab of USTC for providing us with
powerful machines without which this work would not be possible.

M.~D. acknowledges the support of both the Merton College Palmer Scholarship and Oxford Australia Scholarship.
W.~G.~acknowledges the support from the National Science Foundation of China under grant No.~11235010.
YHH would like to thank the Science and Technology Facilities Council, UK, for an Advanced Fellowship and for STFC grant ST/J00037X/1, the Chinese Ministry of Education, for a Chang-Jiang Chair Professorship at NanKai University, the city of Tian-Jin for a Qian-Ren Scholarship, the US NSF for grant CCF-1048082, as well as City University, London, the Department of Theoretical Physics and Merton College, Oxford, for their enduring support.
D.~Z.~acknowledges the support from the National Science Foundation
of China under grant No.~11105138 and 11235010, as well as Zhao-Long Wang and Seung-Joo Lee for their helpful discussions.


\begin{thebibliography}{99}
\bibitem{Gray:2005sr}
J. Gray, Y.-H. He, V. Jejjala, and B.D. Nelson,
``Vacuum geometry and the search for new physics'',
Phys.~Lett.~B638:253--257 (2006).

\bibitem{Gray:2006jb}
J. Gray, Y.-H. He, V. Jejjala, and B.D. Nelson,
``Exploring the vacuum geometry of N=1 gauge theories'',
Nucl.~Phys.~B750:1--27 (2006).

\bibitem{Gray:2008yu}
J. Gray, A. Hanany, Y.-H. He, V. Jejjala, and N. Mekareeya,
``SQCD: A Geometric Apercu'',
JHEP~0805, 099 (2008).

\bibitem{Hanany:2010vu}
A. Hanany, E.E. Jenkins, A.V. Manohar, and G. Torri,
``Hilbert Series for Flavor Invariants of the Standard Model'',
JHEP~1103, 096 (2011).

\bibitem{Katz:1996fh}
  S.~H.~Katz, A.~Klemm and C.~Vafa,
  ``Geometric engineering of quantum field theories''，
  Nucl.~Phys.~B 497, 173 (1997),
  [hep-th/9609239].

\bibitem{sing}
G.-M.~Greuel, G.~Pfister, H.~Sch\"onemann,
  ``Singular: a computer algebra system for polynomial
    computations'', Centre for Computer Algebra, University of
  Kaiserslautern (2001),  Available at {\small\tt http://www.singular.uni-kl.de/}.

\bibitem{m2}
D.~G.~Grayson, M.~E.~Stillman,
``Macaulay2, a software system for research in algebraic geometry'',
available at {\small\tt http://www.math.uiuc.edu/Macaulay2/}.

\bibitem{sage}
  William A. Stein et al.,
  ``Sage Mathematics Software''
  The Sage Development Team, {\small\tt http://www.sagemath.org}.
  For toric CY3, cf.~A.~Novoseltsev and V.~Braun,
  {\small\tt http://www.sagemath.org/doc/reference/schemes/sage/schemes/toric/variety.html}.


\bibitem{Gray:2008zs}
  J.~Gray, Y.~-H.~He, A.~Ilderton, A.~Lukas,
  ``STRINGVACUA: A Mathematica Package for Studying Vacuum Configurations in String Phenomenology,''
  Comput.~Phys.~Commun.~180, 107 (2009),
  [arXiv:0801.1508 [hep-th]].

\bibitem{comp-book}
Y.-H. He, P. Candelas, A. Hanany, A. Lukas and B. Ovrut, Ed,
``Computational Algebraic Geometry in String and Gauge Theory'',
Special Issue, Advances in High Energy Physics, Hindawi
  publishing, 2012, ISBN: 978-0-8218-9136-0.

\bibitem{Mehta:2012wk}
  D.~Mehta, Y.~-H.~He and J.~D.~Hauenstein,
  ``Numerical Algebraic Geometry: A New Perspective on String and Gauge Theories'',
  JHEP~1207, 018 (2012),
  [arXiv:1203.4235 [hep-th]].

\bibitem{Hauenstein:2012xs}
  J.~Hauenstein, Y.~-H.~He and D.~Mehta,
  ``Numerical Analyses on Moduli Space of Vacua'',
  JHEP~1309, 083 (2013),
  [arXiv:1210.6038 [hep-th]].


\bibitem{Cecotti:1992rm}
  S.~Cecotti and C.~Vafa,
  ``On classification of N=2 supersymmetric theories'',
  Commun.~Math.~Phys.~158, 569 (1993),
  [hep-th/9211097].

\bibitem{Cecotti:2011rv}
  S.~Cecotti and C.~Vafa,
  ``Classification of complete N=2 supersymmetric theories in 4 dimensions'',
  Surveys in differential geometry, vol.~18 (2013),
  [arXiv:1103.5832 [hep-th]].

\bibitem{Gaiotto:2009we}
  D.~Gaiotto,
  ``N=2 dualities'',
  JHEP {\bf 1208}, 034 (2012),
  [arXiv:0904.2715 [hep-th]].

\bibitem{He:1999xj}
  Y.~-H.~He,
  ``Some remarks on the finitude of quiver theories'',
  [hep-th/9911114].

\bibitem{Benvenuti:2004dw}
  S.~Benvenuti and A.~Hanany,
  ``New results on superconformal quivers'',
  JHEP~0604, 032 (2006),
  [hep-th/0411262].

\bibitem{Hanany:2012mb}
  A.~Hanany, Y.~-H.~He, C.~Sun and S.~Sypsas,
  ``Superconformal Block Quivers, Duality Trees and Diophantine Equations,''
  JHEP~1311, 017 (2013),
  [arXiv:1211.6111 [hep-th]].

\bibitem{Benvenuti:2006qr}
  S.~Benvenuti, B.~Feng, A.~Hanany and Y.~-H.~He,
  ``Counting BPS Operators in Gauge Theories: Quivers, Syzygies and Plethystics'',
  JHEP~0711, 050 (2007),
  [hep-th/0608050].

\bibitem{Hewlett:2009bx}
  J.~Hewlett and Y.~-H.~He,
  ``Probing the Space of Toric Quiver Theories'',
  JHEP~1003, 007 (2010),
  [arXiv:0909.2879 [hep-th]].

\bibitem{Witten:1993yc}
  E.~Witten,
  ``Phases of N=2 theories in two-dimensions'',
  Nucl.~Phys.~B 403, 159 (1993),
  [hep-th/9301042].

\bibitem{Douglas:1997de}
  M.~R.~Douglas, B.~R.~Greene and D.~R.~Morrison,
  ``Orbifold resolution by D-branes'',
  Nucl.~Phys.~B 506, 84 (1997),
  [hep-th/9704151].

\bibitem{Feng:2000mi}
  B.~Feng, A.~Hanany and Y.~-H.~He,
  ``D-brane gauge theories from toric singularities and toric duality'',
  Nucl.~Phys.~B 595, 165 (2001),
  [hep-th/0003085].

\bibitem{Hanany:2005ve}
  A.~Hanany and K.~D.~Kennaway,
  ``Dimer models and toric diagrams'',
  hep-th/0503149.

\bibitem{Franco:2005sm}
  S.~Franco, A.~Hanany, D.~Martelli, J.~Sparks, D.~Vegh and B.~Wecht,
  ``Gauge theories from toric geometry and brane tilings'',
  JHEP~0601, 128 (2006),
  [hep-th/0505211].

\bibitem{hart}
R.~Hartshorne, ``Algebraic geometry'', GTM 52, Springer-Verlag, 1977.


\bibitem{Douglas:1996sw}
  M.~R.~Douglas and G.~W.~Moore, ``D-branes, quivers, and ALE instantons", hep-th/9603167.

\bibitem{Fulton}
W. Fulton, Introduction to toric varieties, vol. 131 of Annals of
Mathematics Studies. Princeton University Press, Princeton, NJ,
1993. The William H. Roever Lectures in Geometry.

\bibitem{Oda}
T. Oda, Convex bodies and algebraic geometry, vol. 15 of Ergebnisse
der Mathematik und ihrer Grenzgebiete (3) [Results in Mathematics
and Related Areas (3)]. Springer-Verlag, Berlin, 1988. An
introduction to the theory of toric varieties, Translated from the
Japanese.

\bibitem{Leung:1997tw}
  N.~C.~Leung and C.~Vafa,
  ``Branes and toric geometry'',
  Adv.~Theor.~Math.~Phys.~2, 91 (1998),
  [hep-th/9711013].
  
\bibitem{Wess:1992cp}
  J.~Wess and J.~Bagger, ``Supersymmetry and supergravity", Princeton, USA: Univ. Pr. (1992) 259p
  
\bibitem{Luty:1995sd}
  M.~A.~Luty and W.~Taylor,
  ``Varieties of vacua in classical supersymmetric gauge theories",
  Phys.~Rev.~D 53, 3399 (1996),
  [hep-th/9506098].
  
\bibitem{Bhardwaj:2013qia}
  L.~Bhardwaj and Y.~Tachikawa,
  ``Classification of 4d N=2 gauge theories",
  [arXiv:1309.5160 [hep-th]].

\bibitem{Greene:2013ida} 
  B.~Greene, D.~Kagan, A.~Masoumi, D.~Mehta, E.~J.~Weinberg and X.~Xiao,
  ``Tumbling through a landscape: Evidence of instabilities in high-dimensional moduli spaces'',
  Phys.~Rev.~D 88, 026005 (2013),
  [arXiv:1303.4428 [hep-th]].

\bibitem{Aravind:2014}
	Aditya Aravind, Dustin Lorshbough and Sonia Paban,
	``Lower Bound for the Multi-Field Bounce Action'',
	[arXiv:1401.1230 [hep-th]].

\bibitem{savage}
Alistair Savage, Peter Tingley,
``Quiver grassmannians, quiver varieties and the preprojective algebra'',
Pacific J. Math. 251-2 (2011), 393--429,
[arXiv:0909.3746 [math.RT]].

\bibitem{cyril}
	Yang-Hui He, Vishnu Jejjala, Cyril Matti, Brent Nelson,
	``Veronese Geometry of the Electroweak Sector'', to appear.

\bibitem{stanley}
 R.~Stanley,
 ``Hilbert functions of graded algebras'',
 Adv.~Math.~28, 57-83 (1978).

\bibitem{Forcella:2008bb}
  D.~Forcella, A.~Hanany, Y.~-H.~He and A.~Zaffaroni,
  ``The Master Space of N=1 Gauge Theories'',
  JHEP~0808, 012 (2008),
  [arXiv:0801.1585 [hep-th]].

\bibitem{Garey:1979}
  Michael R.~Garey, David S.~Johnson,
  Computers and Intractability: A Guide to the Theory of NP-Completeness, 
  W. H. Freeman and Co., San Francisco, Calif., 1979.

\bibitem{igraph}
  Gábor Csárdi, Tamás Nepusz, The igraph software package for complex network research. InterJournal Complex Systems, 1695, 2006. Avalable at {\small\tt http://igraph.org/}.

\end{thebibliography}
\end{document}